\documentclass{aa}     
\usepackage{graphicx}

\newcommand{\Mpc}{$h^{-1}$\thinspace Mpc}

\def\apj{ApJ}
\def\apjl{ApJL} 
\def\apjs{ApJS}  
\def\aa{A\&A}

\begin{document}


\title{Clusters and Superclusters in the Sloan Digital Sky Survey}

\author {J. Einasto\inst{1},  G. H\"utsi\inst{1},  M. Einasto\inst{1},
 E. Saar\inst{1}, D. L. Tucker\inst{2}, V. M\"uller\inst{3},
 P. Hein\"am\"aki\inst{1,4}, S. S. Allam\inst{5,2} }

\authorrunning{J. Einasto et al.}

\offprints{J. Einasto }

\institute{Tartu Observatory, EE-61602 T\~oravere, Estonia
\and
 Fermi National Accelerator Laboratory, MS 127, PO Box 500, Batavia,
IL 60510, USA
\and
Astrophysical Institute Potsdam, An der Sternwarte 16,
D-14482 Potsdam, Germany
\and
Tuorla Observatory, V\"ais\"al\"antie 20, Piikki\"o, Finland
\and
National Research Institute for Astronomy \& Geophysics, Helwan
Observatory, Cairo, Egypt
}
\date{ Received   09.12.2002 / Accepted ...  }

\titlerunning{SDSS clusters and superclusters}

\abstract{ We apply the 2-dimensional high-resolution density field of
galaxies of the Early Data Release of the Sloan Digital Sky Survey
with a smoothing lengths 0.8~\Mpc\ to extract clusters and groups of
galaxies, and a low-resolution field with smoothing lengths 10~\Mpc\
to extract superclusters of galaxies.  We investigate properties of
density field clusters and superclusters and compare properties of
these clusters and superclusters with Abell clusters, and
superclusters found on the basis of Abell clusters.  We found that
clusters in high-density environment have a luminosity a factor of
$\sim 5$ higher than in low-density environment.  There exists a large
anisotropy between the SDSS Northern and Southern sample in the
properties of clusters and superclusters: most luminous clusters and
superclusters in the Northern sample are a factor of 2 more luminous
than the respective systems in the Southern sample.

\keywords{cosmology: observations -- cosmology: large-scale structure
of the Universe; clusters of galaxies}
}

\maketitle

\section{Introduction}

Clusters and groups of galaxies are the basic building blocks of the
Universe on cosmological scales.  The first catalogues of clusters of
galaxies (Abell \cite{abell}, Zwicky et al. \cite{zwicky}) were
constructed by visual inspection of the Palomar Observatory Sky Survey
plates.  More recent catalogues of clusters, as well as catalogues of
groups of galaxies, have been derived using catalogues of galaxies
(\cite{1983ApJS...52...89H}, \cite{1997MNRAS.289..263D}).  
Moving up the hierarchy of large-scale structure, galaxy cluster
catalogues themselves have been used to define still larger systems
such as superclusters of galaxies (Einasto et
al. \cite{e1994},~\cite{e1997}, ~\cite{e2001}, hereafter E94, E97 and
E01).

The goal of the present paper is to map the Universe up to redshift
$z=0.2$ and to find galaxy clusters and superclusters using the
density field method.  The application of the density field is not
new.  In the pioneering study by \cite{1982ApJ...254..437D} the
density field was used to calculate the gravitational field of the
nearby Universe.  Gott, Melott \& Dickinson ~(\cite{gmd86}) used the
density field to investigate topological properties of the Universe.
\cite{1990ApJ...364..370B} calculated the potential, velocity, and
density fields from redshift-distance data.  Saunders et
al. (\cite{s91}) applied the density field to map the Universe, to
find superclusters and voids, and calculated moments of the density
field. Marinoni et al. (\cite{mar99}) reconstructed real space local
density field. In these studies nearby optical or infrared galaxy
samples were used.  More recently, Hoyle et al. (\cite{h2002}) used
smoothed 2-dimensional density fields from volume limited subsamples
of SDSS EDR galaxies to discuss the 2-dimensional geometry of the
large-scale matter distribution in comparison with $\Lambda$CDM
simulations, and Sheth et al. (\cite{s2002}) advertised a new method
of evaluating isodensity contours of smoothed 3-dimensional density
fields from simulations for characterizing topological properties of
the supercluster-void network.

\begin{table*}
      \caption[]{Data on SDSS EDR galaxies, clusters and superclusters}

         \label{Tab1}
      \[
         \begin{tabular}{cccccccccccc}
            \hline
            \noalign{\smallskip}
            Sample& DEC & RA & $\Delta$RA &
      	$\alpha_1$&$M_1^{\ast}$&$\alpha_2$&$M_2^{\ast}$&
	$N_{\rm gal}$&$N_{\rm cl}$&$N_{\rm ACO}$& $N_{\rm scl}$\\ 

            \noalign{\smallskip}
            \hline
            \noalign{\smallskip}

SDSS.N&
$0^{\circ}$&$190.25^{\circ}$&$90.5^{\circ}$ &
$-1.06$&$-21.55$&$-1.22$&$-20.80$& 15209& 2868&22&24 \\ 
SDSS.S& $0^{\circ}$&$23.25^{\circ}$&$65.5^{\circ}$&
$-1.06$&$-21.40$&$-1.10$&$-20.71$&11882&2287&16&16\\ 
\\
            \noalign{\smallskip}
            \hline
         \end{tabular}
      \]
   \end{table*}

We calculate the density field of the Sloan Digital Sky Survey Early
Data Release (SDSS EDR) by Stoughton et al. (\cite{s02}), as described
by H\"utsi et al. (\cite{hytsi02}, hereafter Paper I), to find
clusters and superclusters of galaxies, and to investigate their
properties.  Clusters of galaxies from the SDSS were extracted
previously by \cite{2002PASJ...54..515G} using the cut and enhance
method; \cite{2002AJ....123...20K} compared various cluster detection
algorithms based on SDSS data.  In this paper we define clusters as
enhancements of the density field and use various smoothing lengths to
separate systems of galaxies of different size and luminosity.  We use
a high-resolution density field to find clusters and groups of
galaxies.  For simplicity, we use the term ``DF-clusters'' for both
groups and clusters found in the high-resolution density field of
galaxies.  Similarly, we use a low-resolution density field to
construct a catalogue of superclusters of galaxies, and denote them as
``DF-superclusters''. DF-clusters and superclusters are defined as
enhancements of the density field, DF-clusters in a fixed volume ($\pm
2.5$~\Mpc\ from the centre), and DF-superclusters as high-density
regions surrounded by a fixed isodensity contour.  In determining
DF-clusters and superclusters we take into account known selection
effects.  We shall investigate some properties of DF-clusters and
superclusters, and study these clusters and superclusters as tracers
of the structure of the local Universe.

This study is of exploratory character to find the potential of SDSS
data to analyse the structure of the Universe both on small and large
scales.  In this stage we use the fact that SDSS EDR covers only
relatively thin slices; thus we calculate the density field in 2
dimensions only.  As more data will be made available we plan to use a
full 3-dimensional data set to detect clusters and superclusters of
galaxies.  In this exploratory stage of the study we will make no
attempt to convert distances of galaxies and systems of galaxies from
redshift space to true space. This correction applies only to
positions of galaxies and clusters, not to their luminosity.  Due to
smoothing of the density field, small-scale corrections (e.g., the
``Finger-of-God'' Effect) are practically flattened out. 

Although there are large-scale corrections due to the apparent
contraction of superclusters in redshift space (the Kaiser Effect,
\cite{1984ApJ...284L...9K}), 
the actual positional shifts of supercluster centres are very small;
furthermore, they do not alter cluster and galaxy positions in
tangential direction.  The only large-scale redshift distortion is the
radial contraction of superclusters, not the number and luminosity of
clusters within superclusters.  Since supercluster shapes are not the
main target of this present study, we may ignore this effect for now.

In Section 2 we give an overview of the observational data. In Section
3 we find DF-clusters and investigate their properties.  Similarly, in
Section 4 we compose a catalogue of DF-superclusters, identify them
with conventional superclusters, and study their properties.  In
Section 5 we continue the study of DF-clusters and superclusters,
derive the luminosity function of DF-clusters, and analyse these
systems as tracers of the large-scale structure of the universe.
Section 6 brings our conclusions.  The three-dimensional distribution
of DF-clusters and superclusters in comparison with Abell clusters and
superclusters is shown on the web-site of Tartu Observatory. The
analysis of the density field of LCRS shall be published by Einasto et
al. (\cite{e02b}). 

\section{Observational data}

\subsection{SDSS Early Data Release }

The SDSS Early Data Release consists of two slices of about 2.5
degrees thickness and $65-90$ degrees width, centred on celestial
equator, one in the Northern and the other in the Southern Galactic
hemisphere (Stoughton et al. \cite{s02}).  This data set contains over
30,000 galaxies with measured redshifts.  We obtained from the SDSS
Catalogue Archive Server angular positions, Petrosian magnitudes, and
other available data for all EDR galaxies. From this general sample we
extracted the Northern and Southern slice samples using following
criteria: redshift interval $1000 \leq cz \leq 60000$~km s$^{-1}$,
Petrosian $r^*$-magnitude interval $13.0 \leq r^* \leq 17.7$, right
ascension and declination interval $140^{\circ} \leq RA \leq
240.0^{\circ}$ and $-1.2^{\circ} \leq DEC \leq 1.2^{\circ}$ for the
Northern slice, and $340^{\circ} \leq RA \leq 60.0^{\circ}$ and
$-1.25^{\circ} \leq DEC \leq 1.25^{\circ}$ for the Southern slice.
The number of galaxies extracted and the mean $RA$ and $DEC$ are given
in Table~\ref{Tab1}. Distances to galaxies and their absolute
magnitudes were calculated as described in Paper I.  In calculating
the distances we used a cosmological model with a matter density of
$\Omega_{\rm m} = 0.3$ and a dark energy density (cosmological
constant) of $\Omega_{\Lambda} = 0.7$ (both in units of the critical
cosmological density).  Throughout this paper, the Hubble Constant is
expressed as usual in units of $H_0 = h~100$ km~s$^{-1}$~Mpc$^{-1}$.
In calculating absolute magnitudes we used K-corrections and
correction for absorption in the Milky Way (for details see Paper I).

\subsection{Abell clusters and superclusters}

We shall use the sample of rich clusters of galaxies by Abell
(\cite{abell}) and Abell et al. (\cite{aco}) (hereafter Abell
clusters), compiled by Andernach \& Tago (\cite{at98}), with redshifts
up to $z=0.13$.  The sample contains 1665 clusters, 1071 of which have
measured redshifts for at least two galaxies.  This sample was
described in detail by E01, where an updated supercluster catalogue of
Abell clusters was presented.  Superclusters were identified using a 
friend-of-friends algorithm with a neighbourhood radius of 24~\Mpc.

\begin{figure}[ht]
\vspace*{7.5cm}
\caption{ Absolute magnitudes of galaxies (black dots), and magnitudes
of the luminosity window, $M_1$ and $M_2$ (coloured dots aligned in 
almost straight lines), for the Northern slice.  }
%

\label{fig:1}
\end{figure}

\section{Density field clusters}

We shall apply the density field to find density enhancements --
DF-clusters.  The procedure consists of several steps: (1) determining
the parameters of the luminosity function needed in the calculating
the weights of galaxies, (2) smoothing of the density field, and (3)
finding clusters (density enhancements) in the field.  Our method of
cluster finding assumes that clusters lie completely within the slice,
and that there are no other clusters with identical $x,y-$coordinates
but shifted vertically.  These assumptions are fulfilled in most
cases.  The extraction of clusters in SDSS survey using full
3-dimensional density field information is planned in the future.

In calculating the density field we make two assumptions.
First, we regard every galaxy as a visible member of a density
enhancement (group or cluster) within the visible range of absolute
magnitudes, $M_1$ and $M_2$, corresponding at the distance of the
galaxy to the observational window of apparent magnitudes. This
assumption is based on observations of nearby galaxies, which indicate 
that there are really almost no isolated galaxies except perhaps very
low-luminosity and diffuse galaxies not represented in surveys like
the SDSS.  Most galaxies belong to poor groups like our own Galaxy, where
one bright galaxy is surrounded by a number of faint satellites.
Further, we assume that the luminosity function derived for a
representative volume can be applied also for individual groups and
galaxies. As we shall see later, this last assumption is not correct, and we
must use various galaxy luminosity functions in order to calculate
density fields suitable for finding DF-clusters and
DF-superclusters. The parameters of the luminosity function (Schechter
\cite{Schechter76}), $\alpha$ and $M^{\ast}$, are given in Table~\ref{Tab1}.

We are interested to find total luminosities of DF-clusters and
superclusters, thus we shall use in the following analysis the
luminosity density rather than the number density.  Under these
assumptions the estimated total luminosity per one visible galaxy is
\begin{equation}
L_{tot} = L_{obs} W_L, 
\label{eq:ldens}
\end{equation}
where $L_{obs}=L_{\odot }10^{0.4\times (M_{\odot }-M)}$ is the
luminosity of the visible galaxy of absolute magnitude $M$, and 
\begin{equation}
W_L =  {\frac{\int_0^\infty L \phi
(L)dL}{\int_{L_1}^{L_2} L \phi (L)dL}} 
\label{eq:weight}
\end{equation}
is the weight (the ratio of expected total luminosity to expected
luminosity in the visibility window). In the last equation
$L_i=L_{\odot }10^{0.4\times (M_{\odot }-M_i)}$ are luminosities of
the observational window corresponding to absolute magnitudes of the
window $M_i$, and $M_{\odot }$ is the absolute magnitude of the Sun. 
This procedure was used by Tucker et al. (\cite{Tucker00}) in the
calculation of total luminosities of groups of galaxies.

\begin{figure*}[ht]
\vspace*{13.0cm}
\caption{ Upper panels show weights as a function of distance.  Grey
  symbols indicate number-density weights, black symbols
  luminous-density weights.  Lower panels plot luminosities of
  galaxies as a function of distance.  Grey circles mark luminosities
  of observed galaxies, black symbols total luminosities, corrected
  for unobservable part of the luminosity range, eq.~(\ref{eq:ldens}).
  Left panels are for luminosity function parameters of set 1, and right
  panels for set 2, which yield better DF-clusters and DF-superclusters,
  respectively.}
%
\label{fig:2}
\end{figure*}

As an example, we plot in Figure~\ref{fig:1} for the Northern slice
the absolute magnitudes of the window, $M_1$ and $M_2$, as well as
observed absolute magnitudes of galaxies, $M_{obs}$.  In upper panels
of Figure~\ref{fig:2} we show luminosity density weights $W_L$ for both
sets of the luminosity function.  For comparison we show also
number-density weights $W_N$, calculated as described in Paper I.
Lower panels of Figure~\ref{fig:2} show luminosities of galaxies
$L_{obs}$, and expected total luminosities $L_{tot}$ calculated with
eq.~(\ref{eq:ldens}).  Luminosities are expressed in units of
$10^{10}$ Solar luminosities.  We see that the number-density weight
$W_N$ rises monotonically with increasing distance from the observer,
whereas the luminosity density weight $W_L$ rises also toward very
small distances. This is due to the influence of bright galaxies
outside the observational window, which are not numerous (see the
weight $W_N$) but are very luminous.  On larger distances the weight
$W_L$ rises again due to the influence of faint galaxies outside the
observational window.  In calculation of the relative density field
only densities in units of the mean density matter, and in this
respect the mean distance dependence of $L_{tot}$ is not very
different from the distance dependence of the number-density
$N_{tot} \propto W_N$.

\begin{figure*}[ht]
\vspace*{20cm} 
\caption{Density field of the SDSS EDR Northern slice, smoothed with
$\sigma = 0.8$~\Mpc, 2~\Mpc, 10~\Mpc, and 16~\Mpc\ dispersion. Upper
panels were calculated for parameter set 1 (better reproduction of
clusters), lower panels for set 2 (better reproduction of
superclusters). In all cases the density field was reduced to a sheet
of constant thickness.}
%
\label{fig:3}
\end{figure*}

\begin{figure*}[hp]
\vspace*{24cm} 
\caption{Density field of the SDSS EDR slices, smoothed with
$\sigma = 0.8$~\Mpc\ dispersion.} 
\label{fig:4}
\end{figure*}

\begin{figure*}[hp]
\vspace*{24cm} 
\caption{Density field of the SDSS EDR slices, smoothed with
$\sigma = 10$~\Mpc\ dispersion. Open circles note positions of Abell
  clusters located within boundaries of slices.} 
\label{fig:5}
\end{figure*}

\begin{figure*}[ht]
\vspace*{15.0cm}
\caption{Luminosities of DF-clusters as a function of distance.   Upper
panels shows the Northern slice, lower panels the Southern slice; left
and right panels represent parameter set 1 and 2, respectively. } 
\includegraphics{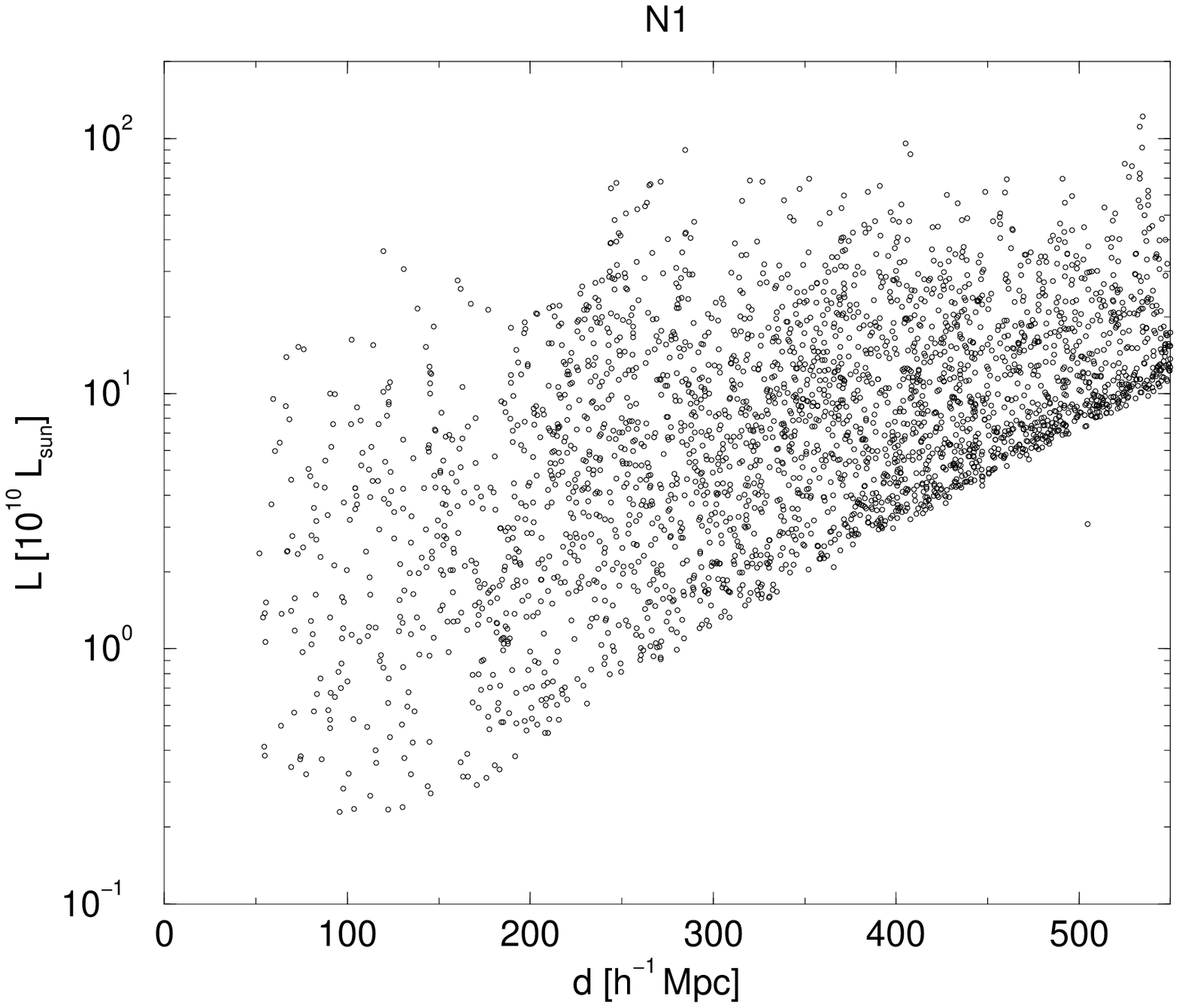}
\includegraphics{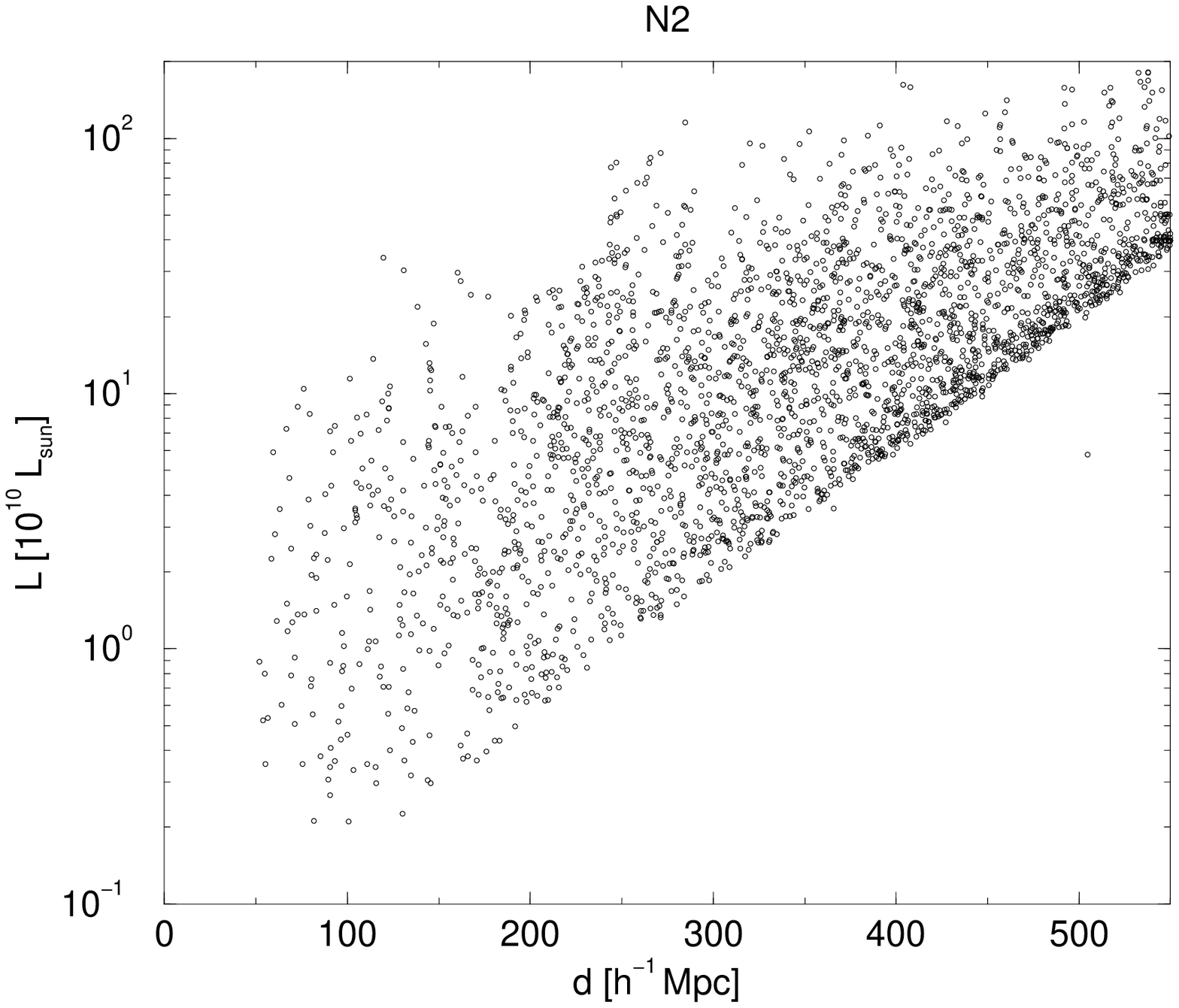}
\includegraphics{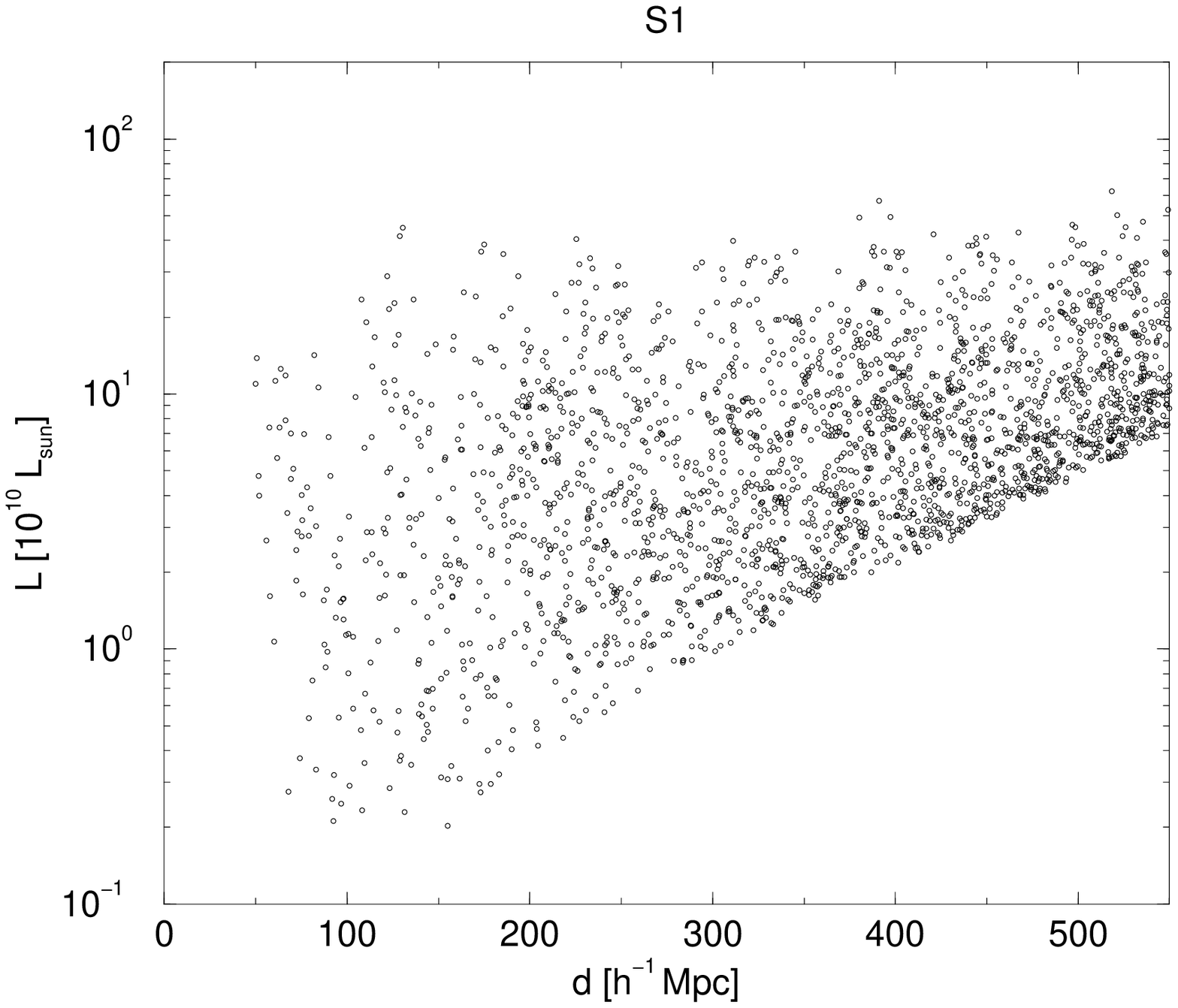}
\includegraphics{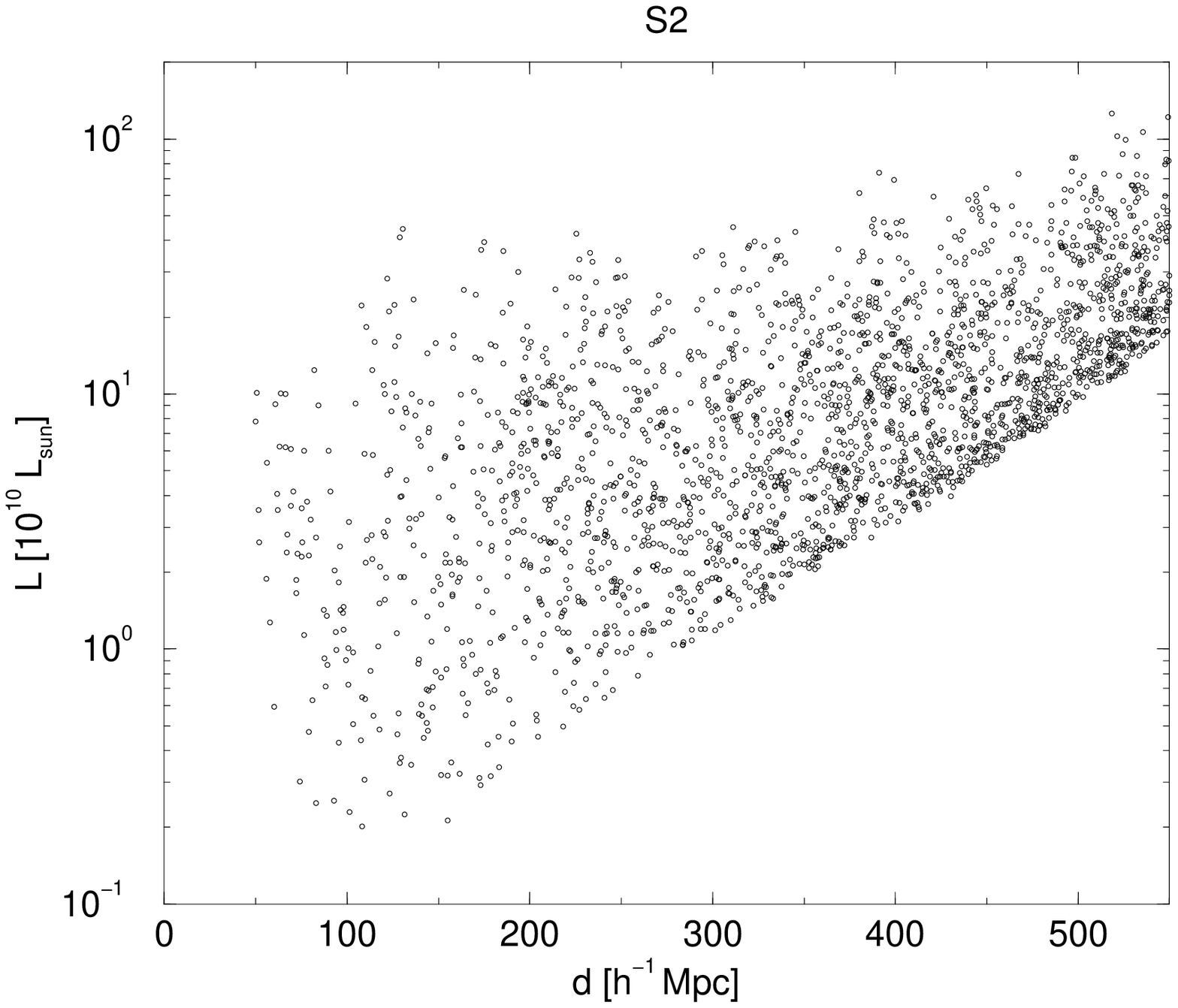}
\label{fig:6}
\end{figure*}

\begin{figure*}[ht]
\vspace*{7.5cm}
\caption{The luminosity density of SDSS EDR Northern and Southern
slices as a function of 
distance, for luminosity function parameters of set 1 (left) and 2 (right). }
\includegraphics{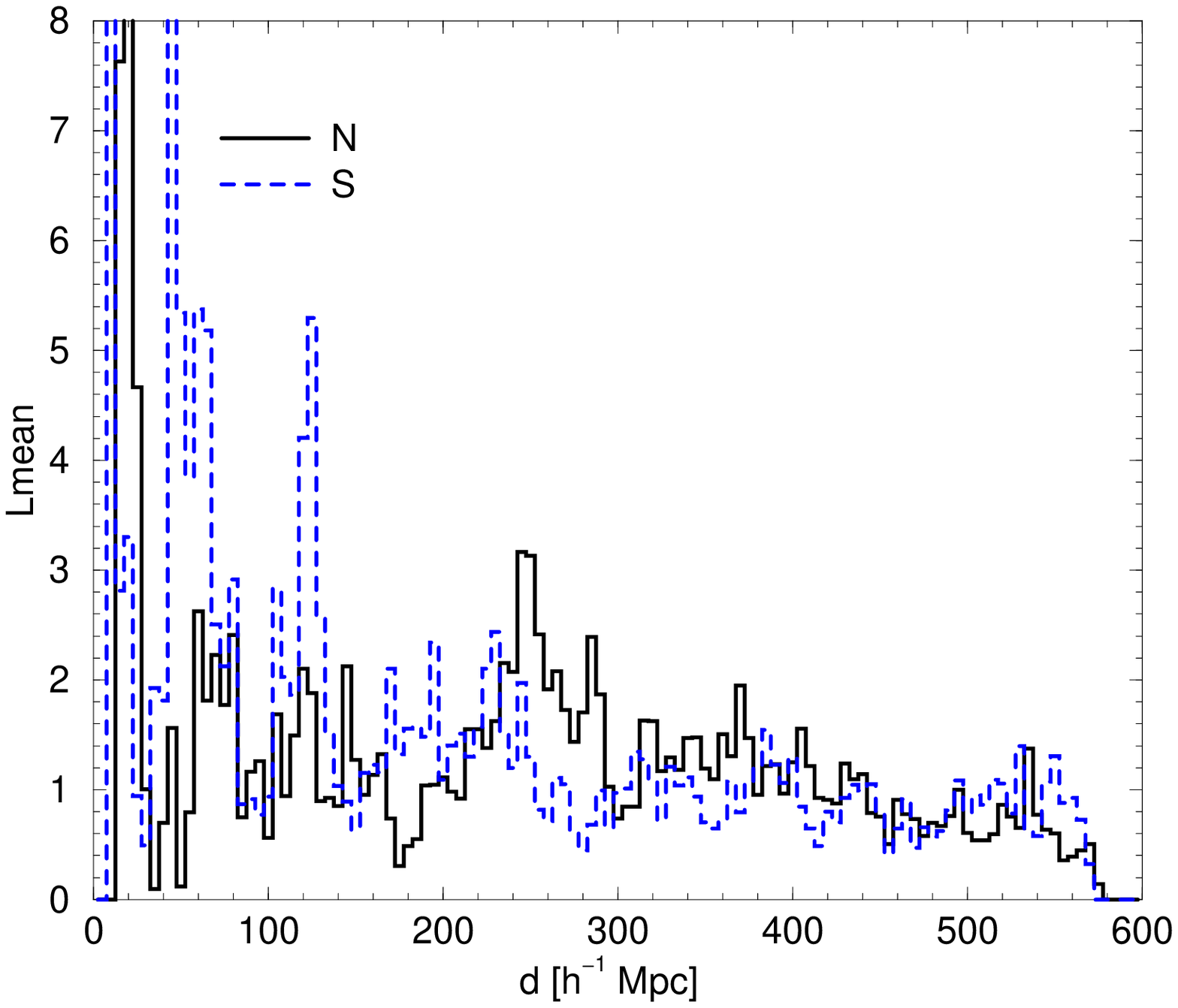}
\includegraphics{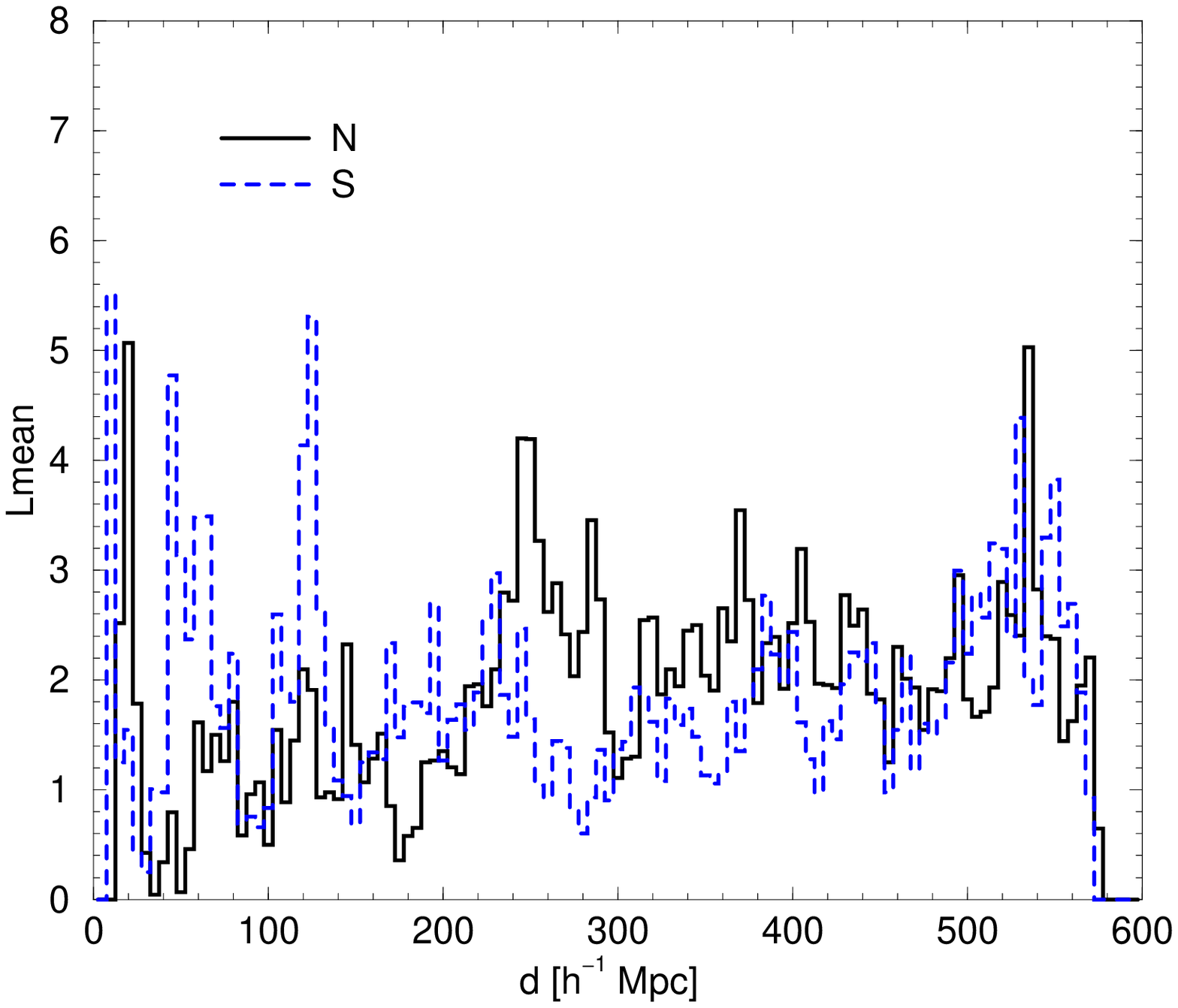}
\label{fig:7}
\end{figure*}

The next step is the smoothing of the field by a Gaussian kernel and
detecting of density field clusters. Here the proper choice of the
smoothing length plays a crucial role. Tests with various smoothing
lengths have shown that with increasing smoothing length we can find
systems of galaxies of increasing scale from clusters and groups and
galaxy and cluster filaments to superclusters and their aggregates.
Figure~\ref{fig:3} shows the density field of the Northern slice
calculated with dispersions $0.8 - 16$~\Mpc. To find DF-clusters with
properties close to those of conventional clusters and groups a
high-resolution density field is to be used with smoothing scale
comparable to the characteristic scale of clusters and groups.  The
harmonic mean radius of groups and clusters in the Las Campanas
Redshift Survey lies in the range $0.45 - 0.65$~\Mpc\ (Tucker et
al. \cite{Tucker00}, Einasto et
al. \cite{e2002}). Abell, APM, and X-ray clusters which have been
matched to Las Campanas clusters have harmonic mean radii in the same
range.  We have used a larger smoothing length, 0.8~\Mpc, to identify
DF-clusters; this smoothing length is equal to the upper quartile of
harmonic radii of groups and clusters (Einasto et al. \cite{e2002}).
With this smoothing length and Gaussian density profile we get density
distributions which match satisfactorily surface density profiles of
clusters of galaxies (\cite{1997ApJ...478..462C}).  Thus we get a
reasonable size of groups even in the case when groups are represented
by only a single bright galaxy (we repeat that we consider every
galaxy as a representative of a group which includes galaxies outside
the observational magnitude window).
Most massive clusters in our sample are represented by a number of
galaxies with a spread in spatial positions.  Thus a modest Gaussian
widening does not considerably distort the actual profile of the
cluster.  After the Gaussian smoothing we get density maps shown in
Figure~\ref{fig:3} --
\ref{fig:5}.

The final step is the identification of the DF-clusters.  To identify
DF-clusters of galaxies, every cell of the field was checked to see
whether the density of the cell exceeded the density of all
neighbouring cells.  If the density of the cell was higher than that
of all its neighbours, then the cell was considered as the centre of a
DF-cluster and its $x, y$-coordinates, density value, and luminosity
were measured. Inspection of the density field has shown that
practically all compact systems of galaxies (groups and clusters)
contain their luminous mass within a box of size $-2 \leq \Delta x
\leq 2$, and $-2 \leq \Delta y \leq 2$ in cell size units.  The
luminosity was derived by adding luminosity densities of the 24
surrounding cells to the luminosity density of the central cell.  All
densities and luminosities were calculated in Solar luminosity units.
We also calculated the density in units of the mean density, this
relative density was used in Figures~\ref{fig:3} -- \ref{fig:5} for
plotting.  A cluster was added to the list of DF-clusters if its
luminosity ${\cal L}$ exceeds a certain threshold value; we used the
threshold ${\cal L}_0 = 0.4 ~10^{10}~ L_{\odot}$.  This threshold
exceeds the luminosity of poorest groups by a factor of about 2, the
number of groups below this limit is very small, see Figure~\ref{fig:6}. 
All used cells were
marked and in the search for new clusters cells already counted were
not included.  Figure~\ref{fig:4} shows that there are only a few
galaxies in the nearby region, $d < 100$~\Mpc.  Moreover, near the
outer boundary of the sample, $d > 550$~\Mpc, samples are already
diluted.  Therefore, we shall confine our DF-cluster catalogue to the
distance interval $100 \leq d \leq 550$~\Mpc.  The total number of
DF-clusters found is given in Table~\ref{Tab1}.

Now we look at the properties of DF-clusters.  Figure~\ref{fig:6}
shows luminosities of DF-clusters as a function of distance from the
observer, $d$.  The lowest luminosity clusters are seen only at a
distance $d \le 150$~\Mpc.  There exists a well-defined lower limit of
cluster luminosities at larger distances, the limit being linear in
the $\log L - d$ plot.  Such behaviour is expected as at large
distances an increasing fraction of clusters does not contain any
galaxies bright enough to fall into the observational window of
absolute magnitudes, $M_1 \dots M_2$.  The cluster lower luminosity limit
is somewhat lower for the Southern slice.  The reason for this
difference is not yet clear.  Nonetheless, the absence of
low-luminosity clusters at large distances can be taken into account
in calculation of the cluster luminosity function (see section 5
below).

According to general cosmological principle the mean density of
luminous matter (smoothed over superclusters and voids) should be the
same everywhere.  Some weak dependence on distance may be due to
evolutionary effects: luminosities of non-interacting galaxies
decrease as stars age; but luminosities of merging galaxies (central
galaxies of clusters) increase with age. Both effects are rather small
within the redshift range $z \leq 0.2$ used in this work
(\cite{2001ApJ...560...72S}). If we ignore these effects we may expect
that the luminous density should not depend on the distance from the
observer, in contrast to the number of galaxies which is strongly
affected by selection effects (at large distances we do not see
low-luminosity galaxies). This difference between the observed and
total luminosity is well seen in Figure~\ref{fig:2}: with increasing
distance the weight $W_L$ (the ratio of the total-to-observed
luminosity) increases by a factor of 3 (parameter set 1) or even 15
(parameter set 2).  We can use the mean total luminous density as a
test of our weighting procedure.  In Figure~\ref{fig:7} we show the
mean luminous density in spherical shells of thickness 5~\Mpc\ for N
and S slices of the SDSS EDR.  We see strong fluctuations of the
luminous density due to superclusters and voids.  The nearby volume is
small, so density fluctuations are large at distances $d \leq
220$~\Mpc; here we also see a low-density region in the Northern slice
(see also Figures~\ref{fig:3} -- \ref{fig:5}). At large distances, the
overall mean density for parameter set 2 is almost constant.  However,
in this case the luminosity of clusters rises considerably with
distance, and all very massive clusters are concentrated near the
outer border of samples (see Figure~\ref{fig:6}).

In determination of the parameter set 2 the number of faint
unobservable galaxies in calculation of the luminosity function was
estimated using Schechter parameters for the whole sample.  As we have
seen above, faint clusters are not visible at large distances.
Thus, in order to estimate correctly the total number of faint galaxies,
the luminous mass of invisible clusters has been added to visible
clusters, and the luminosity of visible clusters has been
overestimated. To get correct luminosities of clusters we have used
the parameter set 1 which yields more or less uniform level of
high-mass clusters at various distances from the observer, see left
panels of Figure~\ref{fig:6}. Schechter parameters of this set
correspond to a subsample of galaxies in high-density environment,
dominating in more distant regions.  Using this set we find, that
luminosities of most massive clusters also increase with
distance. However, a small increase is expected since the volume of
distant shells is larger, and thus the probability to find a bright
galaxy in a shell of fixed thickness is higher on larger distance from
the observer.  Using the luminosity function shown in Paper I we
estimate that most luminous clusters near the outer border of the
sample should be a factor of about 1.25 times more luminous than
clusters in the middle of the sample.  Figure~\ref{fig:6} shows that
this is the case for both the Northern and Southern slice for
parameter set 1.  For this set of parameters the mean luminous matter
density decreases with distance, as seen from Figure~\ref{fig:7}.

The distributions of the mean density and of the cluster luminous
masses are very sensitive tests for parameters of the luminosity
function. Our tests show that it is impossible for one set of
parameters to satisfy both criteria, the distance independence of the
mean luminous density and that of the distribution of luminosities of
clusters.

\section{Density field superclusters}

Superclusters have been traditionally defined either as clusters of
clusters of galaxies (Oort \cite{oort83}, Bahcall \cite{b88}), or as
enhancements in the galaxy distribution (de Vaucouleurs \cite{dev53},
Saunders et al. \cite{s91}, Basilakos, Plionis, \& Rowan-Robinson
\cite{bpr01}, Kolokotronis, Basilakos, Plionis \cite{kbp02}). In the
first case Abell clusters were used to compile supercluster catalogues
(E94, E97, E01).  In the second case superclusters were found by using
the distribution of individual galaxies or by using smoothed density
field maps.  Here we follow the density field approach and use the
low-resolution density field to find large overdensity regions which
we call density field superclusters (or DF-superclusters).  Tests with
various smoothing lengths showed that a smoothing length $\sigma_{sm}
= 10$~\Mpc\ yields a catalogue of superclusters with properties 
similar to those of known superclusters.

The compilation of the supercluster catalogue consists of three steps:
(1) calculating the density field, 2) finding overdensity regions, and
3) determining the properties of the resulting superclusters.  The
density field was calculated as described above with one difference --
in order to reduce the wedge-like volume of slices to a sheet of
uniform thickness we divided densities by the thickness of the slice
at each particular distance.  In this way the surface density of the
field is on average independent of distance, and we can use a distance
independent search for overdensity regions.  The reduced density field
for the Northern and Southern SDSS EDR slices is shown in
Figure~\ref{fig:5}.

\begin{figure}[ht]
\vspace*{13.0cm}
\caption{Properties of SDSS EDR density field superclusters as a function
of the threshold density, $\varrho_0$, to separate superclusters
(high-density regions) and voids (low-density regions). 
The upper panel shows the number of superclusters, $N$, the middle
panel the area of the largest supercluster (in units of the total area
of the slice), and the lower panel -- the diameter (either
in $x$ or $y$ direction, whatever is larger) of the largest
supercluster.} 
\includegraphics{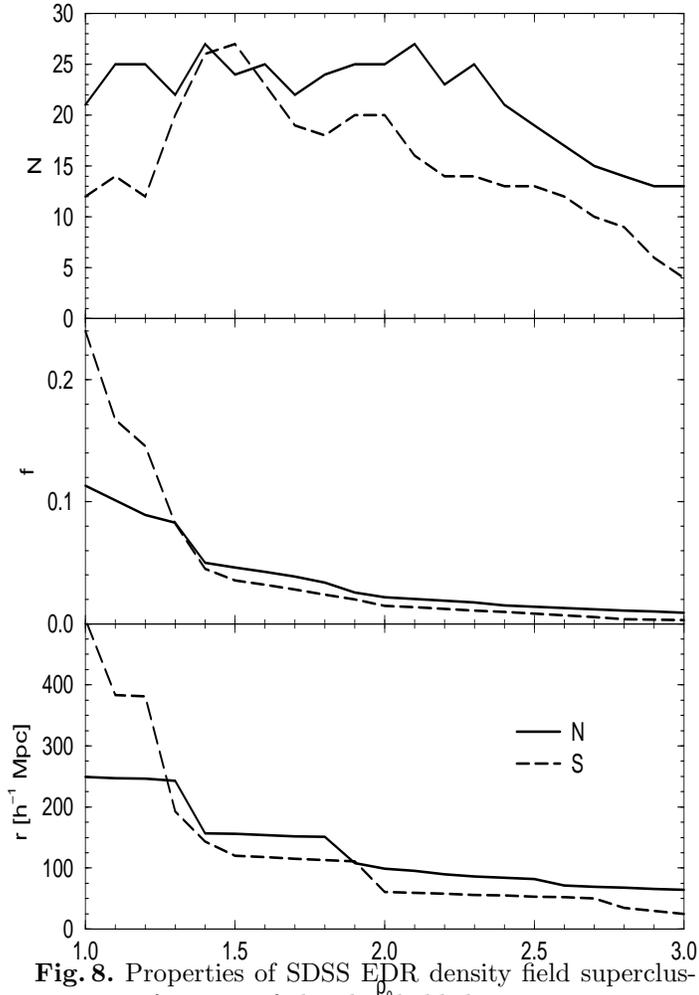}
\label{fig:8}
\end{figure}

Next we searched for connected high-density regions. To do so we need
to fix the threshold density, $\varrho_0$, which divides high- and
low-density regions.  This threshold density plays the same role as
the neighbourhood radius used in the friends-of-friends (FoF) method
to find clusters in galaxy samples and to find superclusters in cluster
samples.  If we have a high threshold density (this corresponds to a
small neighbourhood radius), we get as connected regions only the central
peaks of superclusters -- the number of peaks is small, high-density
regions cover only a small area, and their diameters are small.  When we
decrease the threshold density -- corresponding to an increase in 
the neighbourhood radius in the FoF method --  then the number of
high-density regions increases and their areas and diameters also
increase.  When the threshold density is too low, then superclusters
merge to form large complexes, their masses and diameters grow, and at
a certain threshold density percolation occurs, i.e. the largest
supercluster spans across the whole region under study.

To make a proper choice of the threshold density we plot in
Figure~\ref{fig:8} the number of superclusters, $N$, the area of the
largest supercluster $f$ (in units of the total area of the slice),
and the maximum diameter of the largest supercluster (either in the
$x$ or $y$ direction) as a function of the threshold density
$\varrho_0$ (we use relative densities here). Data are given for both
the Northern and the Southern slices.  We see that the number of
superclusters for the Northern slice has a broad maximum between $1.3
< \varrho_0 < 2.3$, for the Southern slice the maximum is sharper at
$\varrho_0 \sim 1.5$.  For lower $\varrho_0$ large superclusters are
merged; for higher $\varrho_0$ fewer regions are counted as
high-density. The area and diameter of the largest supercluster
continue to drop rapidly until $\varrho_0 \sim 2.0$.  This shows that
various criteria suggest different values for the threshold
density. We have used a threshold density $\varrho_0 = 1.8$.  At this
threshold density the largest supercluster still has a diameter over
100~\Mpc, it has several concentration centres (local density
peaks), and its area forms a large fraction of the total area of the
slice.  The merging of large superclusters can be followed when we
look for sudden changes in the relative area and diameter. For both
slices superclusters are separated at $\varrho_0= 2.1$; this density
value is used to resolve the largest supercluster into
sub-superclusters.  Superclusters were found over the distance interval
$100 \le d \le 555$~\Mpc.  We include only superclusters with areas 
greater than 100~(\Mpc)$^2$; otherwise we get as superclusters tiny
regions with diameters of less than 10~\Mpc.

The number of superclusters in each of the two slices is shown in
Table~\ref{Tab1}.  In Table~\ref{tab:sc} we list for individual
superclusters positional, 
physical, and morphological data.  To aid in identifying these
superclusters, we provide an identification number $No$, the right
ascension RA, the distance $d$, and rectangular coordinates, $x$,
and $y$ (in \Mpc) used in the density field plots.  The column
labelled ``Id'' gives indicates the identification number from E01 of
any matches to previously known superclusters based upon the Abell
cluster sample.

For physical data we present the observed total luminosity of the
supercluster $L_{obs}$ (sum of observed luminosities of DF-clusters
located within the boundaries of superclusters), the estimated total
luminosity of the superclusters $L_{tot}$, the diameter $D$ of the
supercluster (diameter of a circular area equal to the area of the
supercluster), and $\Delta = max(dx,dy)$, where $dx$ and $dy$ are
diameter of the supercluster in the $x$ and $y$ direction.  The
difference between the diameter $D$ and $\Delta$ yields information on
the flattening of the supercluster in the plane of the slice. For
round systems $D \sim \Delta$; for elongated superclusters, $\Delta$
exceeds $D$.  The total luminosity $L_{tot}$ was calculated from the
observed luminosity $L_{obs}$
\begin{equation}
L_{tot} = {D \over D_d} L_{obs},
\end{equation}
where $D_d$ is the thickness of the slice at the distance of the
centre of the supercluster, and we have assumed that the vertical
diameter of the supercluster is identical to the diameter in the plane
of the slice $D$. Finally, $f$ is the fraction of the area of the
supercluster in units of the total area occupied by superclusters in 
a given slice.

For morphological data we give the number of DF-clusters and Abell
clusters lying within the boundaries of the supercluster, $N_{cl}$ and
$N_{ACO}$, respectively, see also Figure~\ref{fig:5}.  Following is
the type of the supercluster ``T'', estimated by visual inspection of
the density field.  Here we have used the following tentative
classification: if the supercluster shows filamentary character, then
its type is ``F'' for a single filament or ``M'' for a system of
multiple filaments.  As an example of a filamentary system we refer to
supercluster N04; an example of a multi-filamentary system is the
supercluster N02.  N04 consists of a single well-defined filament, and
N02 of two massive filaments situated at almost right angles to each
other.  In both cases the main filaments are surrounded by a loose
cloud of faint clusters.  If such diffuse clusters dominate and the
filamentary character is not evident, then the supercluster morphology
is listed as a diffuse ``D'', an example being N08.  Finally, ``C''
denotes a compact supercluster without clear filamentary system, an
example being N15.  This classification is based on the distribution
of galaxies and clusters in a plane.  If 3-dimensional data were
available, then the true shape of each supercluster could be
established.

It should be noted that most superclusters are surrounded by faint
systems of galaxies, either in the form of filaments or of a
diffuse cloud of clusters. Thus the distinction between a filamentary
and diffuse system is not very strict.

\section{DF-clusters and superclusters as traces of the structure of
the universe} 

\subsection{Selection effects}

The main selection effect in the SDSS survey is due to the finite
width of the apparent magnitude window, which excludes galaxies brighter
or fainter than this window.  The presence of a relatively
narrow visibility window affects our results in two different ways.
First of all, it influences the number and luminosity of galaxies in
DF-clusters.  This effect can be taken into account statistically in
the determination of the luminosity of clusters using weighting of
galaxies in DF-clusters. But the visibility window also affects the
number of DF-clusters -- a faint DF-clusters will not be visible at all at 
large distances  if none of its member galaxies is bright enough to
fall within the visibility window.  The number of faint DF-clusters can be
estimated statistically using the procedure described below in 
calculating the DF-cluster luminosity function.  This procedure, however, 
cannot restore physical parameters or positions of individual invisible
DF-clusters, similar to the first procedure which cannot restore
parameters of  invisible galaxies.

The weighting of visible galaxies in clusters statistically restores
properties of clusters which have at least one galaxy in the
visibility window of the survey.  The procedure depends crucially on
the parameters of the galaxy luminosity function which is used to
derive the weights for the visible galaxies.  We have used the
constancy of the mean density with redshift to check the values for
the luminosity function (parameter set 2).  Our analysis has shown
that the mean properties of superclusters are independent of the
distance from the observer, which suggests that the global properties
of the low-resolution density field are correct.  However, in this
case, the luminosities of the DF-clusters are too high at these large
distances.  The reason for the overcorrection of visible cluster
luminosities is simple: visible clusters have to include also
luminosities of invisible clusters. It is clear that in this case we
get a wrong distribution of cluster luminosities.  This is clearly seen 
in Figure~4 of Paper I, where almost all distant clusters seem reddish
(they have very high luminous density), whereas in nearby regions
clusters have various colours from red to yellow, green and blue,
representing clusters of various luminosity.

To avoid this effect we have used another set of values for the
parameters of the galaxy luminosity function (parameter set 1).  Now we
get cluster luminosities which are, in the mean, independent of the
distance from the observer.  In this case the mean density of luminous
matter decreases with distance, as do the mean luminosities of
superclusters.  This deficit of low-luminosity clusters can be
estimated in the determination of the cluster luminosity function (see
next section).

\begin{figure}[hb]
\vspace*{8.0cm}
\caption{Integrated luminosity function of DF-clusters for parameter
sets 1 and 2.   } 
\includegraphics{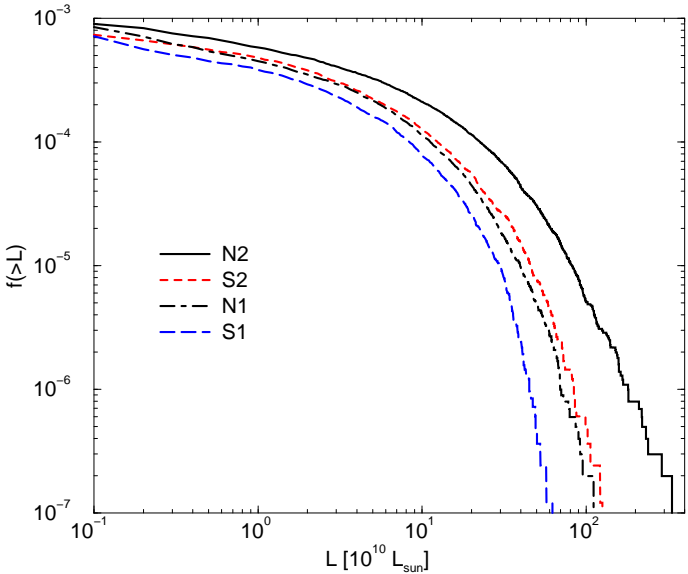}
\label{fig:9}
\end{figure}

There are two possibilities to correct for the second selection effect
in the calculation of the density field.  We can add to the density
field, calculated on the basis of the {\em observed} galaxies and
corrected for unobserved galaxies associated with them, a smooth
background which takes into account unseen galaxies not-associated
with observed ones.  In the calculation of the low-resolution density
field this procedure yields the correct large-scale distribution of
luminous matter -- if the unseen galaxies are distributed randomly.
However, this last assumption needs justification: faint galaxy
systems, like bright systems, may be preferentially located within
superclusters.  On the other hand, this procedure also adds some luminous
matter to DF-clusters, since luminous matter is collected into
DF-clusters from the whole search box $\pm 2.5$~\Mpc\ from the cluster
centre.  In other words, this correction method distorts DF-cluster
properties.  As we want to derive statistically correct cluster
properties, we must use a high-resolution density field without the
background correction.  Taking into account these difficulties we have
used a different method to find the low-resolution density field, by
using for clusters and superclusters various parameters of the galaxy
luminosity function.
We plan to return to this problem when more SDSS galaxy data become
available.

\subsection{Luminosity function of DF-clusters}

Now we calculate the integrated luminosity function of DF-clusters,
i.e. the number of DF-clusters per unit volume exceeding a 
luminosity $L$.  Figure~\ref{fig:6} shows that only very bright
DF-clusters are observable over the whole depth of our samples.  These
bright clusters form a volume limited subsample of DF-clusters. 
Fainter clusters are seen in the nearby volume only; the estimated total
number of fainter clusters can be found by multiplying the observed
number of clusters by a weighting factor of $(d_{lim}/d_L)^3$, where
$d_{lim} = 550$~\Mpc\ is the limiting distance used in compiling of the
DF-cluster sample and $d_L$ is the largest distance where clusters of
luminosity $L$ are seen.  Figure~\ref{fig:6} indicates that there exists an
almost linear relation between $d_L$ and $\log L$; this relation was
used to calculate the weighting factor for clusters of every luminosity.  
Results for both SDSS samples are shown in Figure~\ref{fig:9}.

We see that the luminosity functions span over 3 orders in luminosity and
almost 4 orders in spatial density.  This shows that the SDSS sample
is well suited for determining this function from observations.
Presently we have not determined masses of DF-clusters; thus it is not
possible for us to transform the luminosity function to mass function of
clusters.  However, if we assume that the mass-to-luminosity ratio of
DF-clusters is constant, then we can have an estimate of the
DF-cluster mass function.  Our luminosity function is not very
different from the real observed mass function of conventional
clusters and groups of galaxies (Ramella, Pisani \& Geller \cite{ram},
Girardi \& Giuricin \cite{gir:giu}, Hein\"am\"aki et al. \cite{hei2002}).

The integrated luminosity function was found for both parameters sets
of the galaxy luminosity function (see Figure~\ref{fig:9}).  The
number of bright clusters is much larger for the parameter set 2.  As
discussed above, this is due to the overcorrection of cluster
luminosities -- particularly at large distances -- to compensate for
non-detected faint clusters. Thus we believe that the function for
parameter set 1 corresponds better to reality.

Our results also show that the Northern SDSS sample has a larger
number of luminous DF-clusters than does the Southern sample.  This
difference is seen for both parameter sets of the galaxy luminosity
function.  In order to find the reason for this difference we repeated
our calculations using different values for the parameters of the
galaxy luminosity function for the Northern and Southern samples, as
suggested by the analysis presented in Paper I.  This did not change
the main result -- there exists a considerable difference in the
distribution of luminosities of DF-clusters in the Northern and
Southern samples.  We return to this problem below.

\begin{figure*}[ht]
\vspace*{8.0cm}
\caption{Luminosities of DF-clusters as a function of the global
relative density $\varrho_0$.   Left
panel shows the Northern slice, right panel the Southern slice. } 
\includegraphics{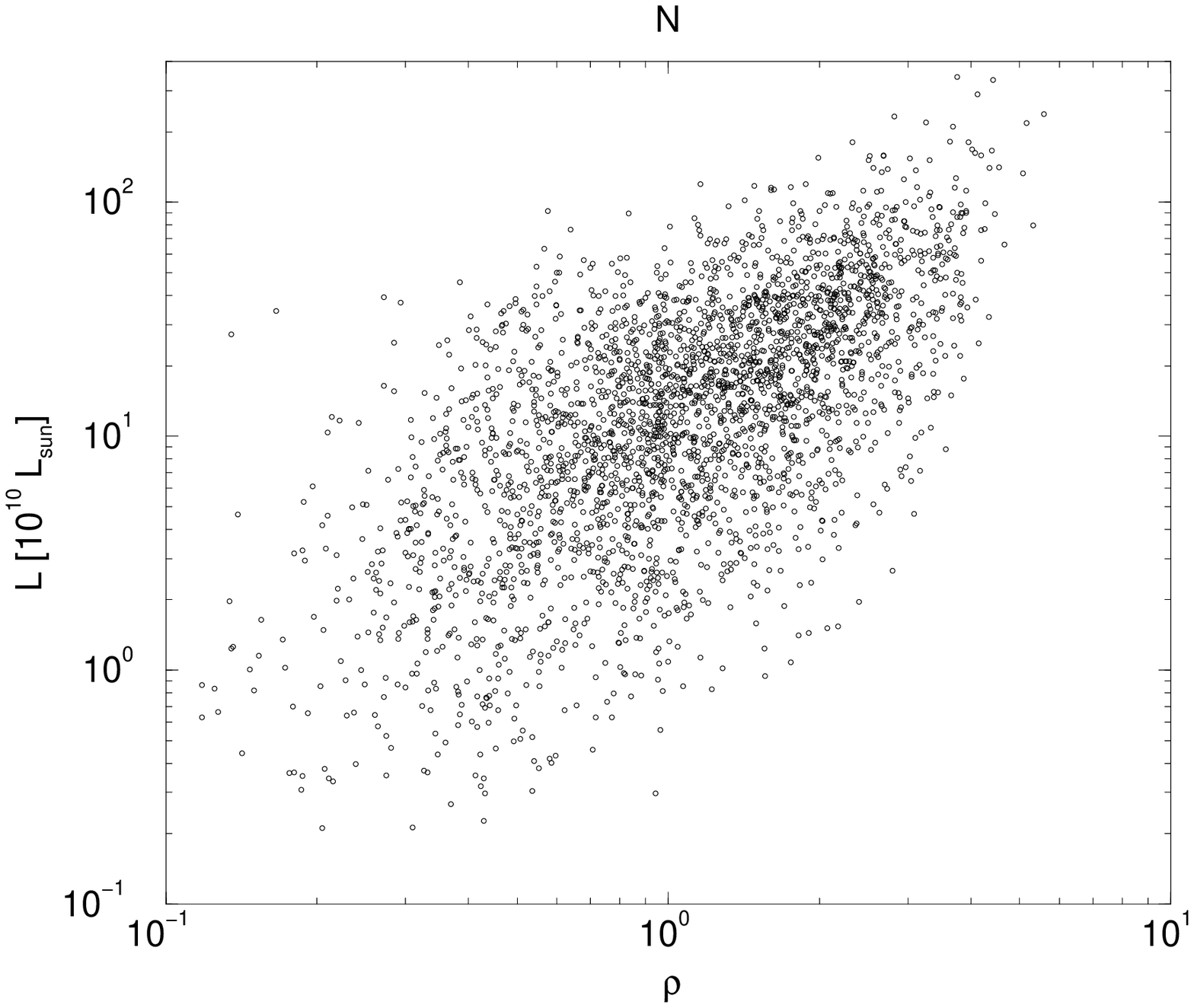}
\includegraphics{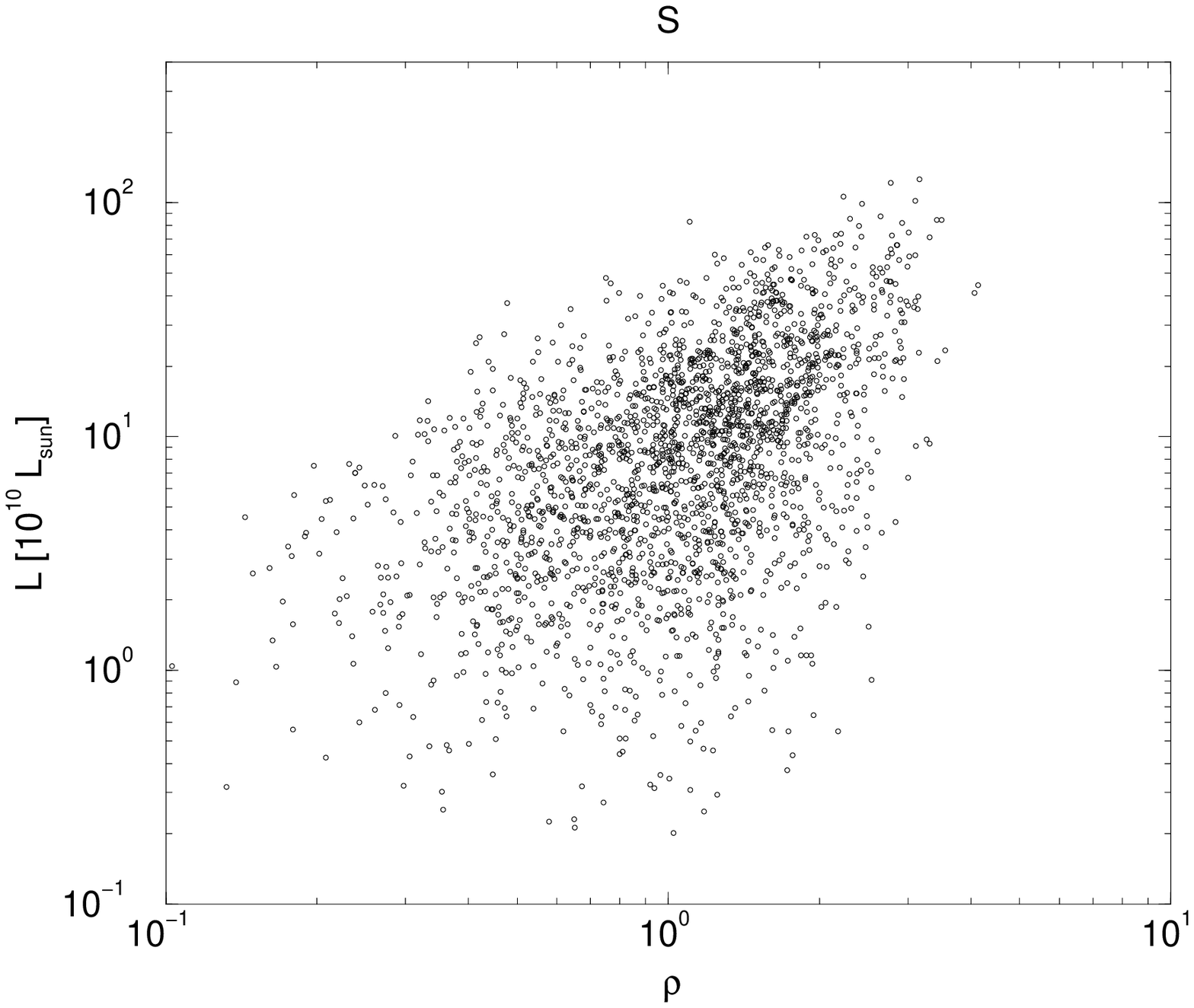}
\label{fig:10}
\end{figure*}

\begin{figure*}[ht]
\vspace*{8.0cm}
\caption{Integrated luminosity functions of DF-clusters as a function
of the global 
relative density $\varrho_0$.   Left
panel shows the Northern slice, right panel the Southern slice. } 
\includegraphics{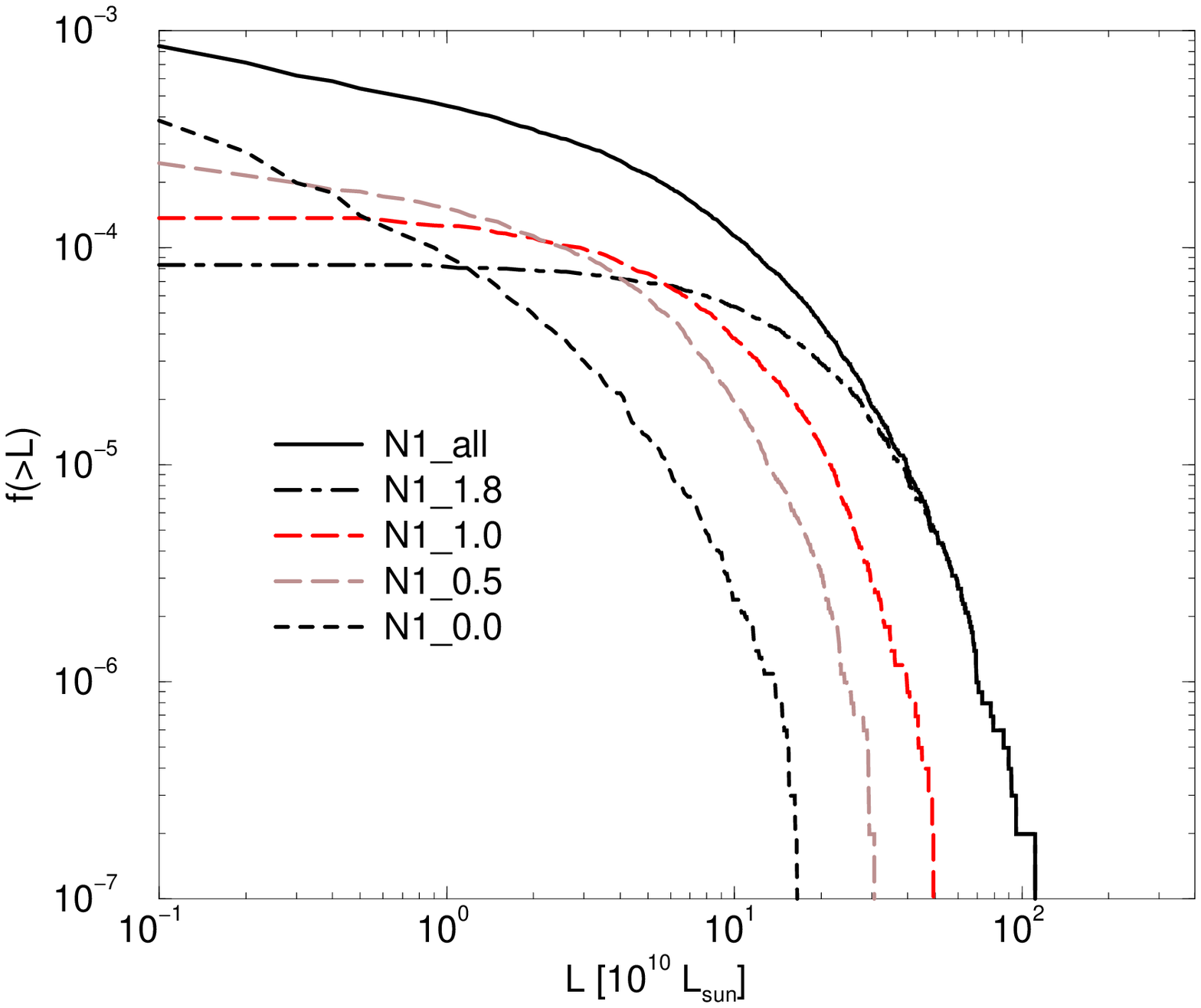}
\includegraphics{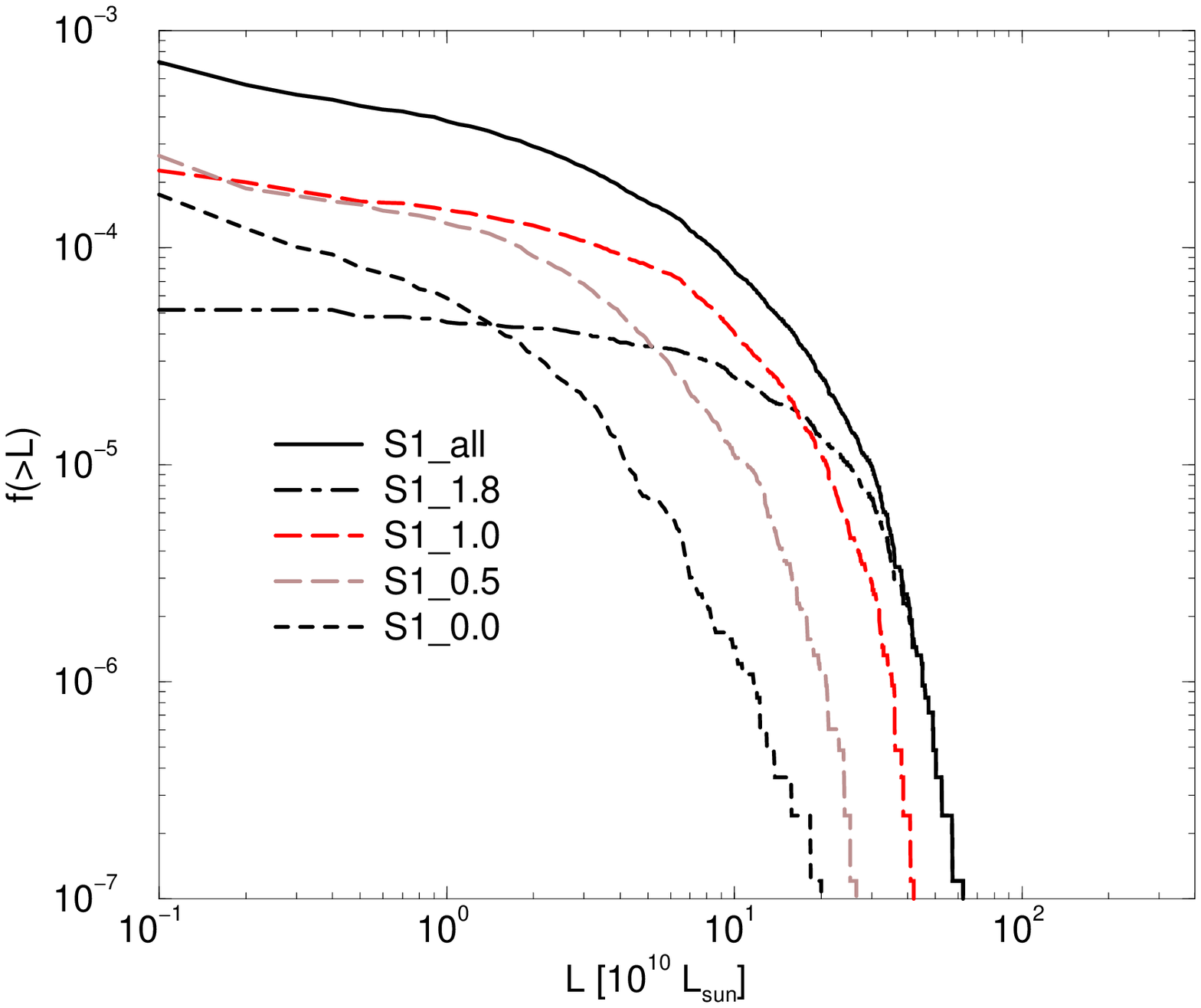}
\label{fig:11}
\end{figure*}

\subsection{Environmental effects in the distribution of DF-clusters}

We can use the density found with the 10~\Mpc\ smoothing as an
environmental parameter to describe the global density in the
supercluster environment of clusters.  We calculated this global
density $\varrho_0$ (in units of the mean density of the
low-resolution density field) for all DF-clusters.  Figure~\ref{fig:10}
shows the luminosity of DF-clusters as a function of the global
density $\varrho_0$.  There is a clear correlation between the
luminosity of DF-clusters and their environmental density.  Luminous
clusters are predominantly located in high-density regions, and
low-luminosity clusters in low-density regions.  This tendency is seen
also visually in Figure~\ref{fig:4}.  Here densities are colour-coded,
we see that small clusters in voids have blue colour which indicates
medium and small densities, whereas rich cluster having red colour
populate dominantly central high-density regions of superclusters.

Figure~\ref{fig:10} shows also that the Northern sample has much more
luminous clusters than the Southern sample.  This result confirms the
presence of a difference between the structure of Northern and
Southern samples.

Figure~\ref{fig:11} shows the cluster luminosity functions of the
Northern and Southern slice calculated separately for 4 global density
intervals, $0 < \varrho_0 \leq 0.5$, $0.5 < \varrho_0 \leq 1.0$, $1.0
< \varrho_0 \leq 1.8$, and $1.8 < \varrho_0 \leq 10$, labelled in
Figure and Table~\ref{tab:dfmed} as N0.0, N0.5, N1.0 and N1.8, for
Northern subsamples, and S0.0, S0.5, S1.0, S1.8 for Southern ones.
The functions were calculated for the luminosity function parameter
set 1 (which yields almost constant highest luminosity DF-clusters for
various distances from the observer).  We list in Table the number of
DF-clusters in subsamples, $N_{\rm group}$, the luminosity of
most-luminous clusters, $L_{\rm lum}$, the mean luminosity of
clusters, $L_m$, and the scatter of luminosities, $\sigma L_m$.  We
see that luminosity functions depend very strongly on the global
environment: the difference in luminosity of most luminous clusters in
subsamples is a factor of $5 \pm 2$.  This difference is not due to
different numbers of clusters, as these numbers are all of the same
order.

One may ask: How much of the cluster luminosity--environment relation
be explained by the fact that a luminous cluster itself contributes to
the luminosity density of the supercluster?  At an extreme, suppose
there is only one cluster in the supercluster.  In that case, the
supercluster's luminosity density is determined entirely by the single
cluster.  In reality, small scatter of the luminosity--environment
relation at high environmental density is probably due just to
proximity to luminous clusters themselves. This influence decreases
when we move toward lower environmental densities.  We plan to
investigate this problem, as well as the influence of selection
effects to the luminosity-density relationship of DF-clusters when
more SDSS data will be available.  The influence of selection is
minimal to luminosities of most-luminous clusters.

\begin{table}
\caption[]{Luminosities of DF-clusters}

\begin{tabular}{lcccc}
\hline 

Sample &$N_{\rm group}$&$L_{\rm lum}$& $L_m$ & $\sigma L_m$ \\
  &    & $10^{10} L_{sun}$      &   $10^{10} L_{sun}$ &   $10^{10} L_{sun}$ \\
\hline 

N0.0 & 438 &  17.9 &  $3.66 $ & $ 2.52$\\
N0.5 & 852 &  33.5 &  $7.20 $ & $  4.09$\\
N1.0 & 872 &  50.7 & $11.30 $ & $  6.28$\\
N1.8 & 716 & 121.5  & $21.72$ & $  12.35$\\
\\
S0.0 & 277 & 20.0 &  $3.57 $ & $ 2.15$ \\
S0.5 & 695 & 26.5 &  $5.60 $ & $  3.30$ \\
S1.0 & 965 & 42.1 &  $9.57 $ & $  5.48$ \\
S1.8 & 341 & 62.3 & $16.93 $ & $  9.84$ \\
\hline

\end{tabular}
\label{tab:dfmed}
\end{table}

\subsection{The fine structure of superclusters and voids}

\begin{figure*}[ht]
\vspace*{8.5cm}
\caption{A view to Abell clusters in equatorial coordinates. Filled
circles show Abell clusters located in superclusters of richness 8 and
more members, open circles mark Abell clusters in less rich
superclusters. Strips near the celestial equator note SDSS slices, the
Northern slice is in the left side. Area surrounded by coloured lines
indicates regions where rich superclusters of the Dominant
Supercluster Plane are located.  Large filled circles near the equator
in the middle of the Northern slice show clusters of supercluster
SCL 126. The Galactic zone of avoidance has in equatorial coordinates a
S-shaped curve. }
%
\label{fig:12}
\end{figure*}

The distribution of DF-clusters within DF-superclusters yields
information on the internal structure of superclusters.  A close
inspection of Figures~\ref{fig:4} and \ref{fig:5} shows, that
superclusters have various internal structures: clusters may form a
single filament, a branching system of filaments, or a more or less
diffuse cloud of clusters.  These morphological properties have been
characterised by types F, M, D, and C in Table~\ref{tab:sc}.  This
Table shows that almost all massive superclusters have morphological
type M, i.e. they are multi-branching.  Faint superclusters do not
have a dominant morphological type: among faint superclusters we find
all morphological types.

In the Northern sample about 25~\%\ of all DF-clusters are located in
superclusters; in the Southern sample this fraction is about 15~\%.
An inspection of the high-resolution density map in Figure~\ref{fig:4}
shows that most DF-clusters outside superclusters also form filaments.
Thus we come to the conclusion that there is no major difference in
the shape of cluster systems in supercluster and in non-supercluster
environments. This similarity of the structure within and outside
superclusters is partly due to our formal procedure of defining
superclusters; actually there is a continuous sequence of structures
from single filaments to multiple filaments and superclusters.  The
difference is mainly in the luminosity of clusters.  This observation
can be interpreted as follows: clusters and cluster filaments in
various environments are formed by similar density perturbations, and
these small-scale perturbations are modulated by large-scale
perturbations which make clusters and their filaments richer in
superclusters and poorer in large voids between superclusters [see
also Frisch et al. (\cite{f95})].

\begin{figure}[ht]
\vspace*{8.0cm}
\caption{Total luminosities of DF-superclusters in SDSS EDR slices
at different distances from the observer.} 
\includegraphics{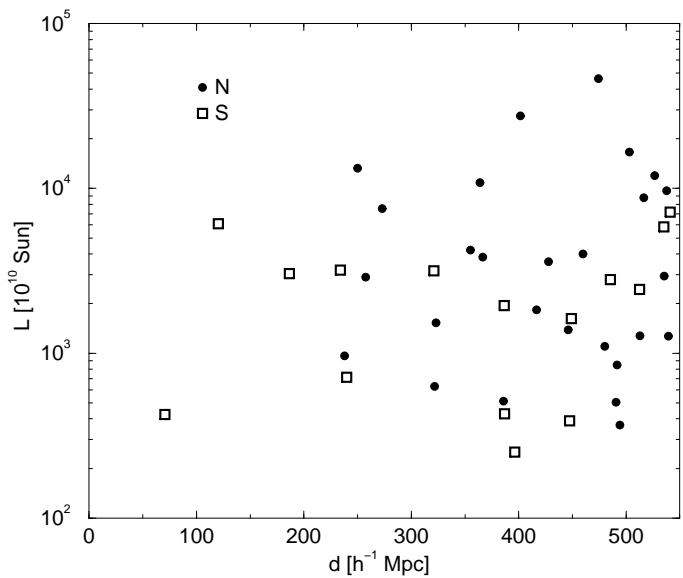}
\label{fig:13}
\end{figure}

Figure~\ref{fig:13} shows the distribution of total luminosities of
DF-superclusters at various distances from the observer.  We see that
both high and low luminosity superclusters are found both in nearby
space as well as in more distant regions.  This distribution is
expected for real superclusters,  and we may expect that there are no
serious systematic errors in the luminosities of  DF-superclusters.

Table~\ref{tab:sc} shows that there are only a few dozen Abell
clusters in EDR SDSS sheets.  In comparison with DF-clusters Abell
clusters form a small minority of clusters, and their distribution
does not match the distribution of even luminous DF-clusters, see
Figure~\ref{fig:5} (we note that in some cases mismatch of Abell and
DF-clusters may be due to distance errors of Abell clusters).  Thus it
is evident that, simply from number statistics alone, Abell clusters
are not as good a structure indicator as are DF-clusters.

\subsection{DF-superclusters in the supercluster-void network}

Now we analyse DF-superclusters as tracers of the large-scale
supercluster-void network.  First we note that most nearby
DF-superclusters can be identified with the Abell supercluster catalogue
of E01, which contains superclusters out to a distance limit of
350~\Mpc\ (see Table~\ref{tab:sc} and Figure~\ref{fig:5}).  Now we
comment on some individual superclusters.

The supercluster N13 (SCL 126 in the list by E01) has the densest
concentration of Abell and X-ray clusters in our neighbourhood.  The
densest part of this supercluster has a declination $-2.5^{\circ}$ and
lies in the $-3^{\circ}$ slice of the Las Campanas Redshift Survey.
It contains 7 Abell clusters and 6 X-ray clusters, most of which are
located within the LCRS slice.  Both in the SDSS and LCRS slices the
supercluster shows a multi-branching filamentary shape.  The form of
the filaments in both slices is different.  This is a good indication
that cluster chains form true filaments, not sheets, since at the
distance of the supercluster both slices are vertically shifted by
only about 12~\Mpc.

Another very rich supercluster is N08 (Virgo-Coma; SCL 111 in E01's
list).  This supercluster contains 15 Abell clusters and 5 X-ray
clusters.  Its centre lies at declination $+9^{\circ}$; thus the SDSS
slice passes through the outskirts of this supercluster. Not far from
E01 SCL 111 lies the supercluster SCL 91 in the list by E97 (Leo-Sextans);
it contains 9 Abell clusters and 1 X-ray cluster at declination
$+3^{\circ}$.  This supercluster is seen as a density enhancement in
Figure~\ref{fig:5}; its density in the SDSS slice is, however, too low
to include this feature as a supercluster in the present catalogue.

There are two more very rich superclusters in the Northern slice:
SCL 107 from E97 (N07) with 8 Abell clusters and 1 X-ray cluster, and
SCL 100 from E97 with 9 Abell clusters.  Their centres lie at
declinations $+10^{\circ}$ and $-3^{\circ}$, respectively, and they
form together with supercluster SCL 111 (N09) a huge complex near the
centre of the SDSS Northern slice.  These superclusters, and
superclusters SCL 82 and SCL 91 belong to the Dominant Supergalactic
Plane -- a collection of very rich superclusters in the nearby
Universe (E97).

The richest supercluster in the Southern slice, according to Abell
supercluster catalogue, is the Pegasus-Pisces SCL 3 from E97  (or S01 in our
catalogue) with 9 Abell clusters, 6 X-ray clusters, and a centre at
declination $+5^{\circ}$.  The second richest supercluster in the
Southern slice is the Pisces SCL 24 (S10) with 7 Abell clusters and 10
X-ray clusters and a centre at declination $+8^{\circ}$.  The SDSS
Southern slice passes through the periphery of these superclusters.

\subsection{The North--South asymmetry in the distribution of clusters and
superclusters}

One of principal results of our study is the discovery of a large
difference between the distribution of galaxies and clusters in the
Northern and Southern slice of the SDSS. A moderate difference is
observed in the galaxy luminosity functions, a much larger difference
exists between distributions of cluster luminosities and between
cluster luminosity functions (see Figure~\ref{fig:9}, ~\ref{fig:10},
and ~\ref{fig:11}).  The North-South asymmetry is seen also as a
systematic difference of total luminosities of Northern and Southern
DF-superclusters: the Northern sample contains more massive
superclusters than the Southern one (see Table~\ref{tab:sc} and
Figure~\ref{fig:13}).  These results indicate the presence of global
differences between the Northern and Southern slices.

One possibility to explain these differences in the distribution of
clusters and superclusters is by various parameters of the galaxy
luminosity function.  To check this possibility we varied parameters
of the luminosity function of galaxies in accordance with results
obtained in Paper I.  In all variants tested so far the North-South
asymmetry remains.  Thus the difference lies probably elsewhere.

Another explanation would be to assume that these differences can be
caused by cosmic variance.  To check this possibility we compared
properties of various systems of galaxies in the Northern and Southern
slices. This has lead us to conclude that the most likely explanation
of the North-South asymmetry is the presence of real differences in
the richnesses of systems of galaxies of various scales.  As we have
discussed in the previous section, the Northern slice lies in a region
of space containing many rich superclusters.  The $\pm 10^{\circ}$
zone around the celestial equator in the right ascension interval of
the Northern slice contains 25 Abell superclusters in the catalogue by
E01 (15 in the $\pm 6^{\circ}$ zone).  Among these, there are 5 very
rich superclusters containing 7 or more Abell clusters (all in the
central $\pm 6^{\circ}$ zone).  The $\pm 10^{\circ}$ zone around the
Southern slice has 12 Abell superclusters; the $\pm 6^{\circ}$ zone
has 7 Abell superclusters, among which there is only one very rich
system.  The distribution of Abell clusters on the celestial sphere is
presented in Figure~\ref{fig:12}.  Regions of the SDSS EDR are marked
by solid strips, dashed lines show regions where rich superclusters
belonging to the Dominant Supergalactic Plane (DSP) are projected on
the sky. The DSP was detected by E97. In this plane are concentrated
numerous very rich superclusters. We see that the Northern SDSS EDR
slice crosses the DSP. Near the supercluster SCL 126 (or N13 in our
list) the DSP crosses another region of rich superclusters seen in the
Northern hemisphere extending from a declination of $+60^{\circ}$ to a
declination of $-30^{\circ}$.  So the Northern slice lies at
crossroads of two complexes of very rich superclusters.  In contrast,
the Southern slice is located in a region away from  very rich
superclusters:  in the Southern Hemisphere, rich superclusters, some of 
which belong to the DSP, are located at lower declinations.  

Thus our tentative conclusion is that the North-South asymmetry is not
a mere random fluke but a real difference in the large-scale
distribution of matter.  So far this difference is manifested in only
a single pair of slices.  This result must be checked by future SDSS
data.

\subsection{Error analysis}

There are three main error sources in our quantitative data: errors
due to number statistics (Poisson errors), errors in the values for
the parameters of the galaxy luminosity function, and errors in our
assumptions concerning the constancy of the mean luminous density and
the upper end of luminosities of DF-clusters. Our quantitative
analysis is related directly to luminosities of DF-clusters. Thus we
must analyse how these errors affect cluster luminosities.

The simplest errors are related to the Poisson statistics.  In Paper I
we determined Poisson errors for parameters of the luminosity function
of galaxies.  It is easy to estimate the influence of these errors to
properties of density field clusters.  We have repeated the density
field analysis with many sets of parameters.  We show here results for
a third set of parameters $M^{\star} = -20.73$ and $\alpha = -1.14$,
the difference between sets 2 and 3 being $\Delta M^{\star} = 0.07$, and
$\Delta \alpha = 0.08$, which corresponds approximately to the $2\sigma$
error of parameters in the luminosity function (see Paper I).  We have
found DF-clusters for both parameter sets and made
cross-identification of clusters.  Cluster positions are practically
identical, only total luminosities differ.  Both luminosities are
compared in Figure~\ref{fig:14}.  We see that there exists a very
close relationship with small scatter between luminosities of
DF-clusters found with parameter sets 2 and 3.  Our main quantitative
results (North-South asymmetry of luminosities and dependence of
luminosities on environment) are differential, thus errors of
parameters of the luminosity function of galaxies do not alter these
conclusions.

\begin{figure}[ht]
\vspace*{8.0cm}
\caption{Comparison of luminosities of DF-clusters for different
parameter sets. In vertical axis we plot luminosities of clusters for
parameter set 2, in horizontal axis either for set 1 or 3.} 
\label{fig:14}
\end{figure}

A much more serious systematic difference is due to our poor
understanding of the global behaviour of the mean luminosity density,
i.e. the third error mentioned above.  The influence of this error is
also demonstrated in Figure~\ref{fig:14}, where we compare
luminosities of DF-clusters for parameter sets 1 and 2.  Here the
scatter is much larger -- almost all clusters are much brighter for
the parameter set 2.  This is the reason why cluster luminosity
functions for parameter sets 1 and 2 differ so much.  The physical
reason for this large uncertainty is our poor understanding of the
general behaviour of the density field -- how constant it should be
for different distances from the observer.  We plan to come back to
this problem in a future study.

There are other sources of uncertainty, ones which influence positions
of DF-clusters and superclusters.  These errors are due to the fact
that in this paper we have used positions of galaxies in redshift
space.  Redshift space distortions are of two different types: they
elongate clusters of galaxies in the radial direction (finger-of-god
effect), and they shift positions of DF-clusters and superclusters in
the radial direction toward the contraction centre (bulk motions).  As
superclusters are more massive attractors than clusters, their
positions are much less influenced than are positions of clusters.  Due to
bulk motions supercluster shapes are slightly compressed in the radial
direction.  The shift is small and does not influence luminosities of
clusters and superclusters.

The largest uncertainty in the properties of superclusters is due to
the fact that we presently have information only for thin slices.  In
their largest dimension, a supercluster may extend for hundreds of
Mpc; this is much, much larger than the smallest dimension of our
slices, their thickness.  Therefore, it is impossible to say for
certain just what fraction of a given supercluster lies within a
slice's boundaries.  Our estimate of the total luminosity of
superclusters is based on the assumption that they have in vertical
direction a structure similar to the structure on the plane observed.
This assumption is correct if the observed plane passes close to the
centre of the supercluster.  If the observed plane passes through the
periphery of the supercluster, we get an underestimated value of the
total luminosity and diameter.  Thus values shown in
Table~\ref{tab:sc} can be considered as lower limits of actual values
of the respective quantities.

One possible error of cluster luminosities is due to possible
overlapping of clusters in $x, ~y-$coordinates due to our
2-dimensional detection scheme.  The probability of this error is,
however, rather low.  In the future we plan to define clusters using
full 3-dimensional data where this error is excluded.

\section{Discussion and conclusions}

Results of this paper and Paper I are of a methodical and 
quantitative character.  Methodical aspects concern the application of
the density field method with various smoothing lengths to display and
describe systems of galaxies of various scales.  We calculated density
field for a number of smoothing lengths from 0.8 to 16~\Mpc. In this
way systems of galaxies on various scales could be visualised and their
mutual relationship could be studied.  The high-resolution density field, 
with dispersion 0.8~\Mpc\, was used to define density field clusters
and to investigate the structure of superclusters.  The low-resolution
density field, with dispersion 10~\Mpc\, was used to define
superclusters of galaxies and to study global properties of clusters
and superclusters.

The inspection of the distribution of clusters in superclusters and
voids suggests that galaxy systems have in both regions similar shape
in that the dominant structural elements are single or multi-branching
filaments.  Massive superclusters have dominantly a multi-branching
morphology; less massive superclusters have various morphologies,
including compact, filamentary, multi-branching, and diffuse systems.

Quantitative results concern the luminosity function of galaxies and
properties of clusters and superclusters.  We have found in Paper I
and confirmed in this paper that it is impossible to find a global set
of parameters of the luminosity function which can be applied in all
cases.  We found that parameters of the luminosity function are
different for high- and low-density regions: galaxies in high-density
regions are more luminous.  This result confirms earlier findings by
Lindner et al. (\cite{l95}): bright galaxies define larger voids than
faint ones. More recently this conclusion was obtained by
\cite{1998ApJ...505...25B}, \cite{2000ApJ...545....6B}, and by
\cite{2001MNRAS.328...64N}.  The present study indicates that the
effect is larger than previously suspected: luminosities of the brightest
galaxies in high-density regions exceed  the luminosities of the brightest
galaxies in low-density regions by a factor of about 5.

The values of the parameters of the luminosity function are also
different for nearby and distant parts of the survey and for the
Northern and Southern slices.  A distance dependence of parameters of
the luminosity function has been found for deeper surveys spanning
redshift interval up to $z=1$ by \cite{1999ApJ...518..533L} and by
\cite{2001ApJ...560...72S} in the Canadian Network for Observational
Cosmology Redshift Survey.  However, in this survey the distance
dependence appears only by comparison of nearby ($z \sim0$) and
distant ($z > 0.2$) parts of the survey.  Thus it seems improbable
that this effect can explain our results on the distance dependence in
a relatively small redshift interval $0 < z \leq 0.2$.

Parameters of the luminosity function depend also on the density of
the environment.  This dependence influences properties of the density
field. High-density regions contain brighter galaxies than do
low-density ones.  This difference leads to different selection effects 
for galaxies in high- and low-density environments.  In a high-density
environment, due to selection effects, faint galaxies in clusters
cannot be observed, but clusters themselves are visible (since they
contain at least one galaxy bright enough) and the total
luminosity of clusters can be estimated to take into account
unobserved galaxies.  In a low-density environment all galaxies of the
cluster may be too faint, and the cluster may not be seen at all.  
In another words, distance-dependent selection effects influence
clusters and superclusters in different ways.

The density field was calculated using the expected total luminosities of
clusters of galaxies, including the expected luminosities of galaxies too
faint or too bright to be included in the redshift survey.  The
correction for unobserved galaxies was made assuming a Schechter
luminosity function for galaxies. Our analysis shows that it is
impossible to correct the density field so that general properties of
the density field and properties of clusters of galaxies are correct
for the same set of parameters of the galaxy luminosity function. 
For this reason we have used two sets of parameters of the
luminosity function. Parameter set 1 has a strong bright-end
(parameter $M^{\star} = -21.55$) and corresponds to galaxies in a
high-density environment which dominate clusters observed at large
distance; this set yields correct properties of clusters of galaxies,
but does not include faint clusters at large distances, and thus gives
too low of a density for distant superclusters. Parameter set 2 was obtained
for the whole region under study, and it has moderate bright-end
(parameter $M^{\star} = -20.80$). In this parameter set ALL faint
invisible galaxies at large distance are included within visible clusters,
including galaxies that belong to non-detected clusters; thus clusters
themselves become too luminous with increasing distance, but
supercluster properties are correct.

Comparing clusters in different environments we have found that there
exists a strong dependence of cluster properties on the density of the
large-scale environment: clusters located in high-density environments 
are a factor of $5 \pm 2$ more luminous than clusters in low-density
environments.

Finally we found that there exists a large difference between
properties of clusters and superclusters in the Northern and Southern
slices of the SDSS EDR survey: clusters and superclusters in the
Northern slice are more luminous than those in the Southern slice by a
factor of 2.  This difference may be due to differences in the
location of slices with respect to the very large-scale environment.
If this conclusion is confirmed by future observations one must
conclude that the formation and evolution of galaxies and systems of
galaxies of various scales depends on the nearby as well as on the
large-scale environment.  Richer superclusters have more luminous
galaxies and clusters.  On smaller scales this tendency has been
observed as a difference between properties of clusters within
superclusters and in voids.  Now we see that a similar difference may
be observed on much larger scales.

\begin{acknowledgements}

The present study was supported by Estonian Science Foundation grants
ETF 2625, ETF 3601, and ETF 4695 and TO 0060058S98.  P.H. was
supported by the Finnish Academy of Sciences.  J.E. thanks Fermi-lab
and Astrophysikalisches Institut Potsdam for hospitality where part of
this study was performed.
\\
Funding for the creation and distribution of the SDSS Archive has been
provided by the Alfred P. Sloan Foundation, the Participating
Institutions, the National Aeronautics and Space Administration, the
National Science Foundation, the U.S. Department of Energy, the
Japanese Monbukagakusho, and the Max Planck Society. The SDSS Web site
is http://www.sdss.org/.
\\
The SDSS is managed by the Astrophysical Research Consortium (ARC) for
the Participating Institutions. The Participating Institutions are The
University of Chicago, Fermi-lab, the Institute for Advanced Study, the
Japan Participation Group, The Johns Hopkins University, Los Alamos
National Laboratory, the Max-Planck-Institute for Astronomy (MPIA),
the Max-Planck-Institute for Astrophysics (MPA), New Mexico State
University, University of Pittsburgh, Princeton University, the United
States Naval Observatory, and the University of Washington.

\end{acknowledgements}

{
\scriptsize
\vskip-2.5cm 
\begin{table*}[ht] 
\begin{center} 
\caption{The list of superclusters} 
\begin{tabular}{rrrrrrrrrcrccc} 
\\ 
\hline 
\\ 
$No$ & $L_{obs}$             & $L_{tot}$             & $D$    & $\Delta$  & RA  & $d$    & $x$    & $y$   
 & $f$ & $N_{cl}$ &    $N_{ACO}$ & Id & T \\   
     & [$10^{10}~L_{\odot}$] & [$10^{10}~L_{\odot}]$ & Mpc
 & Mpc    & deg & 
  Mpc   &   Mpc   &  Mpc  &     &          &         &    &   \\
\\
\hline
\\
S01    &   186.6 &   714.1 &    24.0 &    35.0 &     1.2 &   239.7 &    90.5 &   222.0 &  0.0205 &   11 & 0 & 3& M  \\  
S02    &   554.2 &  1621.4 &    34.4 &    42.0 &     4.8 &   449.0 &   143.5 &   425.5 &  0.0420 &   20 & 0 & &  M \\  
S03    &   345.6 &  3040.3 &    43.0 &    53.0 &     8.8 &   186.5 &    47.0 &   180.5 &  0.0655 &   30 & 2 & 19& M  \\  
S04    &   588.1 &  3164.9 &    45.2 &    63.0 &     9.2 &   320.9 &    79.0 &   311.0 &  0.0726 &   33 & 3 & 14& M  \\  
S05    &   283.7 &   642.6 &    32.8 &    36.0 &     9.7 &   552.4 &   131.5 &   536.5 &  0.0381 &    7 & 0 & &  F \\  
S06    &   159.7 &   251.8 &    16.4 &    23.0 &    13.2 &   396.2 &    70.0 &   390.0 &  0.0095 &    4 & 0 & &  F \\  
S07    &    59.8 &   121.7 &    29.6 &    33.0 &    15.1 &   554.8 &    80.0 &   549.0 &  0.0310 &    2 & 0 & &  C \\  
S08    &   851.2 &  2441.7 &    38.5 &    45.0 &    19.2 &   512.4 &    37.5 &   511.0 &  0.0526 &   22 & 0 & &  M \\  
S09    &   449.8 &  3187.9 &    43.4 &    59.0 &    23.9 &   234.0 &    -2.0 &   234.0 &  0.0670 &   33 & 1 & 34& M  \\  
S10    &   379.0 &  6103.8 &    50.8 &    92.0 &    28.0 &   120.4 &    -9.5 &   120.0 &  0.0915 &   29 & 1 & 24& M  \\  
S11    &   204.5 &   429.3 &    21.3 &    25.0 &    28.9 &   386.8 &   -37.0 &   385.0 &  0.0161 &    5 & 0 & &  M \\  
S12    &   955.7 &  2798.4 &    37.2 &    56.0 &    31.5 &   485.3 &   -68.0 &   480.5 &  0.0492 &   24 & 0 & &  M \\  
S13    &  1496.1 &  5852.5 &    54.8 &    65.0 &    43.2 &   535.0 &  -181.0 &   503.5 &  0.1066 &   35 & 0 & &  M \\  
S14    &   283.9 &   389.6 &    16.1 &    17.0 &    45.9 &   447.5 &  -171.0 &   413.5 &  0.0092 &    7 & 0 & &  M \\  
S15    &   518.9 &  1943.0 &    37.9 &    44.0 &    52.1 &   386.4 &  -185.5 &   339.0 &  0.0510 &   24 & 0 & &  M \\  
\\
N01    &  2340.4 &  8754.9 &    50.6 &    85.0 &   150.6 &   516.4 &   331.5 &   396.0 &  0.0547 &   32 & 0 & &  M \\  
N02    &  1027.0 &  7528.6 &    52.4 &    71.0 &   153.9 &   273.0 &   163.0 &   219.0 &  0.0587 &   36 & 4 & 82 & M  \\  
N03    &   709.3 &  1271.4 &    25.3 &    32.0 &   158.6 &   539.3 &   285.5 &   457.5 &  0.0137 &   11 & 0 & & F  \\  
N04    &   875.1 &  3836.5 &    42.1 &    67.0 &   160.4 &   366.5 &   184.0 &   317.0 &  0.0379 &   26 & 0 & 252& F  \\  
N05    &  1281.3 &  2943.2 &    32.2 &    36.0 &   166.9 &   535.4 &   214.5 &   490.5 &  0.0222 &   13 & 0 & & M  \\  
N06    &   267.4 &   512.4 &    19.3 &    22.0 &   172.9 &   385.7 &   117.0 &   367.5 &  0.0080 &    6 & 0 & & C  \\  
N07    &   228.8 &   628.6 &    23.1 &    26.0 &   178.6 &   321.5 &    66.5 &   314.5 &  0.0114 &    9 & 0 & 107& F  \\  
N08    &   223.5 &   963.8 &    26.9 &    30.0 &   180.8 &   238.0 &    40.5 &   234.5 &  0.0154 &   11 & 0 & 111& D  \\  
N09    &  1833.0 & 10859.9 &    56.4 &    84.0 &   182.3 &   363.8 &    52.5 &   360.0 &  0.0681 &   64 & 2 & 265& M  \\  
N10    &  6412.0 & 46426.6 &    89.9 &   151.0 &   184.0 &   474.1 &    54.0 &   471.0 &  0.1728 &  116 & 0 & & M  \\  
N10A   &   701.1 &  1832.6 &    28.5 &    34.0 &   190.1 &   416.5 &     3.5 &   416.5 &  0.0257 &   20 & 0 & & M  \\  
N10B   &  1288.8 &  3994.9 &    37.3 &    40.0 &   185.6 &   459.7 &    39.5 &   458.0 &  0.0441 &   21 & 0 & & M  \\  
N10C   &  3616.2 & 16608.7 &    60.5 &    95.0 &   181.7 &   503.0 &    77.5 &   497.0 &  0.1158 &   54 & 0 & & M  \\  
N11    &   306.5 &   367.4 &    15.5 &    16.0 &   185.8 &   494.2 &    41.5 &   492.5 &  0.0052 &    5 & 0 & & D  \\  
N12    &   393.5 &   504.7 &    16.5 &    19.0 &   190.7 &   490.5 &    -1.0 &   490.5 &  0.0058 &    7 & 0 & & D  \\  
N13    &  1446.1 & 13240.7 &    59.9 &    90.0 &   198.3 &   249.8 &   -33.5 &   247.5 &  0.0767 &   56 & 1 & 126& M  \\  
N14    &  2704.0 & 11938.6 &    60.9 &    78.0 &   198.5 &   526.5 &   -72.5 &   521.5 &  0.0793 &   50 & 0 & & D  \\  
N15    &   428.0 &  1533.3 &    30.3 &    32.0 &   200.2 &   323.0 &   -54.0 &   318.5 &  0.0196 &   12 & 0 & & C  \\  
N16    &  1115.7 &  3590.8 &    36.0 &    40.0 &   207.3 &   427.7 &  -123.5 &   409.5 &  0.0278 &   25 & 0 & & D  \\  
N17    &   571.3 &  1101.2 &    24.2 &    25.0 &   209.9 &   480.1 &  -159.0 &   453.0 &  0.0126 &   13 & 0 & & D  \\  
N18    &  2532.5 &  9667.2 &    53.7 &    69.0 &   212.1 &   537.6 &  -197.5 &   500.0 &  0.0618 &   34 & 0 & & M  \\  
N19    &   443.6 &   847.8 &    24.6 &    28.0 &   215.3 &   491.5 &  -205.5 &   446.5 &  0.0129 &    8 & 0 & & C  \\  
N20    &  3738.6 & 27566.5 &    77.5 &    94.0 &   216.8 &   401.4 &  -177.5 &   360.0 &  0.1285 &  102 & 2 & & M  \\  
N21    &   656.5 &  1274.0 &    26.1 &    33.0 &   220.1 &   512.8 &  -253.0 &   446.0 &  0.0145 &   10 & 0 & & C  \\  
N22    &   536.3 &  2891.5 &    36.4 &    57.0 &   230.0 &   257.6 &  -163.5 &   199.0 &  0.0283 &   16 & 1 & 153& M  \\  
N23    &   928.3 &  4221.6 &    42.3 &    52.0 &   230.4 &   355.0 &  -227.5 &   272.5 &  0.0382 &   32 & 1 & 155& M  \\  
N24    &   595.3 &  1387.1 &    27.2 &    28.0 &   233.1 &   445.9 &  -301.5 &   328.5 &  0.0158 &   13 & 0 & & M  \\  
\\
S16    &  1572.5 &  7159.3 &    64.5 &   113.0 &   359.1 &   541.1 &   223.0 &   493.0 &  0.1478 &   45 & 0 & & M  \\  
\\
\hline
\label{tab:sc}
\end{tabular}
\end{center}
Distance, sizes and coordinates are given in \Mpc.
\end{table*}
}
\end{document}